\documentclass{emulateapj}

\usepackage{graphicx}
\usepackage[hypertex]{hyperref}

\def \nh {${\rm N_{H}}$}
\def \eg {e.g.}
\def \ie {i.e.}
\def \etc {etc.}
\def\spose#1{\hbox to 0pt{#1\hss}}
\def\ltsim{$\mathrel{\spose{\lower 3pt\hbox{$\sim$}}
        \raise 2.0pt\hbox{$<$}}$\thinspace}
\def\gtsim{$\mathrel{\spose{\lower 3pt\hbox{$\sim$}}
        \raise 2.0pt\hbox{$>$}}$\thinspace}

\def\plotone#1{\includegraphics[width=7in]{#1}}

\newcommand{\thin }{\thinspace}

\newcommand{\lcdm}{$\Lambda$CDM}
\newcommand{\lk }{${\rm L_K}$}

\newcommand{\msun }{${\rm M_{\odot}}$}
\newcommand{\lsun }{${\rm L_{\odot}}$}
\newcommand{\msunlsun}{${\rm M_\odot L_\odot^{-1}}$}
\newcommand{\ergps }{${\rm erg\ s^{-1}}$}
\newcommand{\mvir}{${\rm M_{vir}}$}

\newcommand{\reff}{${\rm R_{e}}$}

\newcommand{\zfe }{${\rm Z_{Fe}}$}

\newcommand{\chandra }{{\em Chandra}}
\newcommand{\genx}{{\em Generation-X}}
\newcommand{\ixo}{{\em IXO}}

\newcommand{\astroh}{{{\em Astro-H}}}

\newcommand{\xspec }{{\em Xspec}}

\newcommand{\hst}{HST}
\newcommand{\minuit}{MINUIT}
\newcommand{\ciao }{\href{http://cxc.harvard.edu/ciao/}{\em CIAO}}
\newcommand{\caldb }{\href{http://cxc.harvard.edu/ciao/}{\em Caldb}}
\newcommand{\heasoft }{\href{http://heasarc.gsfc.nasa.gov/docs/software/lheasoft/}{\em Heasoft}}
\newcommand{\ned}{\href{http://nedwww.ipac.caltech.edu/}{\em{NED}}}

\newcommand{\mbh} {${\rm M_{BH}}$}
\newcommand{\mstars}{${\rm M_*}$}

\newcommand{\fb}{${\rm f_b}$}
\newcommand{\lj }{${\rm L_J}$}

\newcommand{\leda}{{\em{HyperLEDA}}}

\newcommand{\rhog}{${\rm \rho_g}$}

\newcommand{\twomass}{2MASS}
\newcommand{\sigmac}{$\sigma_*$}

\newcommand{\mbondi}{${\rm \dot{M}_{bondi}}$}

\defcitealias{humphrey06a}{H06}
\defcitealias{humphrey08a}{H08}

\slugcomment{Accepted for publication in the Astrophysical Journal}
\begin{document}
\title{Hydrostatic Gas Constraints on Supermassive Black Hole Masses:
Implications for Hydrostatic Equilibrium and Dynamical Modelling
in a Sample of  Early-Type Galaxies}
\author{Philip J. Humphrey\altaffilmark{1}, David A. Buote\altaffilmark{1},
Fabrizio Brighenti\altaffilmark{2,3}, Karl Gebhardt\altaffilmark{4} and William G. Mathews\altaffilmark{3}}
\altaffiltext{1}{Department of Physics and Astronomy, University of California at Irvine, 4129 Frederick Reines Hall, Irvine, CA 92697-4575}
\altaffiltext{2}{Dipartimento di Astronomia, Universit\`{a} di Bologna, Via Ranzani 1, Bologna 40127, Italy}\altaffiltext{3}{University of California Observatories, Lick Observatory, University of California at Santa Cruz, Santa Cruz, CA 95064}
\altaffiltext{4}{Astronomy Department, University of Texas, Austin, TX 78712}
\begin{abstract}
We present new mass measurements for the supermassive 
black holes (SMBHs) in the centres of three early-type galaxies.
The gas pressure in the surrounding, hot interstellar medium (ISM)
is measured through spatially resolved spectroscopy with the 
\chandra\ X-ray observatory, allowing the SMBH mass (\mbh) to be inferred
directly under the hydrostatic approximation. This technique does
not require calibration against other SMBH measurement methods
and its accuracy depends only on the ISM being close to hydrostatic,
which is supported by the smooth X-ray isophotes of the galaxies.
Combined with results from our  recent study of the elliptical galaxy
 NGC\thin 4649, this
brings to four the number of galaxies with SMBHs measured in this 
way. Of these, three already have mass determinations
from the kinematics of either the stars or a central gas disc,
and hence join only a handful of galaxies with \mbh\ measured by 
more than one technique.
We find good agreement between the different methods,
{providing support for the assumptions implicit in both the hydrostatic and 
the dynamical models}.
The stellar mass-to-light ratios for each galaxy 
inferred by our technique are in agreement with the 
{predictions of stellar population synthesis models assuming a \citeauthor{kroupa01a} 
initial mass function (IMF). This concurrence implies that no more than
$\sim$10--20\%\ of the ISM pressure is nonthermal, unless there is a 
conspiracy between the shape of the IMF and nonthermal
pressure. Finally, we compute Bondi accretion rates (\mbondi),
finding that the two galaxies with the highest \mbondi\ exhibit little evidence of 
X-ray cavities, suggesting that the correlation with the AGN 
jet power takes time to be established.}
\end{abstract}
\keywords{Xrays: galaxies--- galaxies: elliptical and lenticular, cD--- galaxies: ISM---galaxies: individual (NGC1332, NGC4261, NGC4472, NGC4649)--- black hole physics}
\section{Introduction} \label{sect_introduction}
Supermassive black holes (SMBHs) with masses ranging from less than
a million to a 
few billion times the mass of the Sun, believed to be ubiquitous in
galaxy bulges, are increasingly being recognizing as essential ingredients
in the formation of galaxies \citep[\eg][]{silk98a,dimatteo05a,hopkins06a}. 
Compelling evidence for the close link between the formation of 
SMBHs and stellar spheroids is provided by the tight
correlations observed between the SMBH mass, \mbh, and the
properties of the galaxy, such as the bulge luminosity, central
light concentration and, in particular, central velocity dispersion,
\sigmac\ \citep{gebhardt00b,ferrarese00b,kormendy95a,marconi03a,graham01a}.
Since different physical processes involved in this co-evolution
can lead to quantitatively different behaviour,
both the shape of these relations,
particularly that between
\mbh\ and \sigmac,
 and the scatter about them
have become crucial diagnostics for constraining our picture
of galaxy formation 
\citep[\eg][]{king03a,nipoti03a,granato04a,robertson06a}.
Unfortunately, controversies persist over the exact slope of the 
\mbh-\sigmac\ relation \citep{merritt01a,tremaine02a,ferrarese05a},
and a number of authors have discussed possible
deviations from the  universal, log-linear form
typically assumed \citep{wyithe06a,lauer07a,hu08a,graham08a}.
{Furthermore, most authors neglect systematic uncertainties in the modelling
techniques and these could have an important effect \citep{gultekin09a}.}

Despite their central importance for understanding 
galaxy evolution, only $\sim$40--50 precise, reliable SMBH measurements
span $\sim$3 orders of magnitude in \mbh\ \citep{ferrarese05a,graham08a,gultekin09a}.
These are all derived from the kinematics of  stars or a rotating
central gas disc in the deep potential well near to the black hole.
In external galaxies, kinematical information is mostly restricted to the 
distribution of velocities along the line of sight, 
 so the use of either tracer
{implies understanding the three-dimensional velocity field. Gas disc
 modeling generally assumes a circular velocity field, so that
deviations from Keplerian motion can give rise to errors in \mbh\
\citep[\eg][]{cappellari02a}.
Stellar kinematics are complicated
by the degeneracy between the enclosed mass and the stellar orbital
structure. To overcome this degeneracy, it is standard to use the
full velocity profile, by either including higher order moments 
\citep{vandermarel93a,bender94a} or using non-parametric fitting
\citep{gebhardt00a}.
More importantly though, it is essential to minimize assumptions about the
stellar orbital structure. For an axisymmetric system, this generally requires 
orbit-based models \citep[\eg][]{gebhardt03a,vandermarel98a,vandenbosch08a}.
However, these models are not free of systematic concerns,
such as whether to regularize \citep{valluri04a,vandermarel98a} or not
\citep[\eg][]{gebhardt03a}, ensuring adequate phase space coverage 
\citep{valluri04a} and the possible implications of triaxiality
\citep[\eg][]{vandenbosch08a}. Since the inclination of an elliptical
galaxy is generally poorly known this can introduce significant uncertainties
into the assumed three-dimensional light distribution, and consequently 
the inferred gravitating mass \citep[\eg][]{gavazzi05a}.
A reliable recovery of \mbh\ also depends
on the accurate modelling of the total gravitating mass profile over a 
wide radial range, so that the results can be sensitive to the treatment
of the dark matter component. Unfortunately, most published SMBH measurements
from stellar dynamics omit this mass component, leading to a systematic 
underestimate of \mbh\ by as much as a factor $\sim$2 \citep{gebhardt09b}.}

Clearly further progress demands
independent verification of the different mass determination techniques.
An effective way to address this question is
to compare multiple \mbh\ measurements {\em for the same SMBH} 
made in different ways. However, reliable measurements of the same 
black hole by both dynamical methods are rare 
{and comparisons between
them yield mixed results. To date
the most
detailed examples are for NGC\thin 4258 \citep{siopsis09a,miyoshi95a}, for which
the measurements were marginally discrepant (at $2.6 \sigma$, but within
$\sim$15\%\ of each other), Cen~A,
for which both techniques agree within a factor $\sim$4 but with
systematic discrepancies between different
measurements (even those made using the same method) that far exceed the
quoted statistical errors \citep{silge05a,marconi06a,neumayer07a,cappellari09a},
and NGC\thin 3379, for which there is general agreement for axisymmetric 
models \citep{shapiro06a,gebhardt00a}, but not with triaxial models
\citep{vandenbosch09a}}
A handful of similar
comparisons are less compelling due to deviations from Keplerian
gas motion, or overly simple stellar modelling
\citep[\eg][]{cappellari02a,verdoeskleijn02a}.

A new way of detecting SMBHs, appropriate for early-type galaxies,
was proposed by \citet{brighenti99c}, who {pointed out that the 
gravitational influence of a quiescent SMBH on the hot, X-ray emitting
interstellar medium (ISM) should affect the temperature and density of 
the gas sufficiently to be detectable in \chandra\ observations of 
nearby galaxies. Although their models predict the ISM to be globally 
inflowing}, the inflow is expected to be highly subsonic so that the ISM
should remain very close to hydrostatic equilibrium,
enabling \mbh\ to be measured directly with the standard
hydrostatic X-ray mass determination methods
\citep[\eg][hereafter \citetalias{humphrey06a}]{gastaldello07a,humphrey06a}.
{If the inflow is adiabatic over scales which can be resolved with \chandra,
the resulting gas compression should produce 
a measurable central temperature spike
near to the black hole.}
Unlike other techniques that do not rely on dynamical modelling,
such as using the variability or luminosity of an active galactic nucleus
\citep[][and references therein]{markowitz03a,oniell05a,ferrarese05a},
hydrostatic \mbh\ measurements {\em do not} require calibration against 
other methods and their intrinsic accuracy on a case-by-case basis
is limited theoretically only by the extent to which the hydrostatic approximation
holds (in addition to further simplifying assumptions usually 
adopted to make the problem more tractable, and provided there is 
sufficient spatial resolution to resolve the gas properties close to the SMBH). 

Since the local freefall timescales in the centres of early-type
galaxies are generally much shorter than the gas cooling time,
hydrostatic equilibrium is expected to be quickly established
unless the ISM is being actively stirred up,
for example by AGN activity or galaxy merging \citep{mathews03a}.
Nonthermal pressure due  to gas turbulence is not expected 
to perturb this equilibrium substantially; 
hydrodynamical simulations of structure formation suggest it 
contributes no more than $\sim$25\%\ of the total pressure
 \citep{evrard96a,nagai07a} and possibly significantly
less \citep{fang09a}. This is consistent with the general
agreement between the masses of morphologically relaxed galaxy
clusters measured by X-ray and gravitational lensing methods
\citep[\eg][]{allen98a,mahdavi08a}. \citet{churazov08a} compared
stellar dynamics and X-ray mass measurements in two early-type
galaxies with moderately disturbed X-ray isophotes and, similarly,
found that nonthermal support contributed no more than $\sim$10--20\%\
of the global pressure.  
The excellent agreement between the
distribution of dark matter in galaxies, groups and clusters 
inferred from X-ray studies and that predicted by cosmological simulations
provides further corroboration of the hydrostatic approximation
(\citealt{lewis03a}; \citealt{vikhlinin06b}; \citetalias{humphrey06a}; \citealt{buote07a}).
Nevertheless, the important question of whether the gas in the galaxies 
{\em studied in this 
paper} is close to hydrostatic is one we will discuss in detail in \S~\ref{sect_he}.

In a recent
\chandra\ study of the giant elliptical galaxy NGC\thin 4649 
\citep[][hereafter \citetalias{humphrey08a}]{humphrey08a}, we used
this technique for the first time to constrain the mass of the SMBH.
This galaxy has an extremely round,
relaxed X-ray morphology, indicating little or no large-scale
gas disturbance, and suggesting little turbulence that would
perturb hydrostatic equilibrium by no more than a few percent 
\citep{brighenti09a}. Using the X-ray data, 
we constrained \mbh\ to 
$\sim 3\times 10^9$\msun, which is close to the 
measurement from stellar dynamics \citep{gebhardt03a}.
This agreement provides compelling evidence not only that the gas must
be close to hydrostatic, but also that the stellar modelling of this
system is accurate, effectively ruling out pathological orbital
structure.

In this paper, we extend our study to include three more
objects, bringing to a total of four  the number of early-type
galaxies in which \mbh\ has been determined from hydrostatic 
X-ray methods. Of these, three already have \mbh\ determined by
stellar dynamics or gas kinematics while, for the remaining 
galaxy, we are presenting the first direct \mbh\ measurement.
In \S~\ref{sect_targets} we discuss the sample selection, 
in \S~\ref{sect_analysis} we describe the X-ray data reduction, in
\S~\ref{sect_stars} we discuss the deprojection of the stellar 
light, an essential step in modelling the mass,
in \S~\ref{sect_models} we outline how we infer the mass from
the X-ray data and in \S~\ref{sect_results} we report the best-fitting
results. We discuss possible sources of systematic uncertainty
in \S~\ref{sect_systematics}  and reach our conclusions in
\S~\ref{sect_discussion}.
{All errors quoted in the text of this paper, unless otherwise stated, correspond
to the 90\%\ confidence regions (which, for our Bayesian analysis, implies the 
region of marginalized parameter space within which the integrated probability
is 90\%). In the figures, we typically use 1--$\sigma$ errors (68\%\ enclosed
probability), for clarity}.

\section{Target selection} \label{sect_targets}
\begin{deluxetable*}{llllllllr}
\tabletypesize{\scriptsize}
\tablecaption{The galaxy sample\label{table_sample}}
\tablehead{
	\colhead{Galaxy}  & \colhead{Distance} & \colhead{\lj} & \colhead{\sigmac} & \colhead{Age}
& \colhead{$<$[Z/H]$>$} & \colhead{ObsID} & \colhead{Date} & \colhead{Exposure} \\
\colhead{} & \colhead{(Mpc)}& \colhead{($10^{11}$\lsun)} & \colhead{(${\rm km\ s^{-1}}$)}& \colhead{(Gyr)} &
\colhead{} & \colhead{} & \colhead{} & \colhead{(ks)} 
}
\startdata
NGC\thin 1332 & 21.3 & 0.86 & 321$\pm$14 & $4.1^{+8.8}_{-1.4}$ & 0.32$\pm0.30$ & 4372 & Sep 19 2002 & 41 \\
NGC\thin 4261 & 29.3 & 1.4  & 308$\pm$6& 15$\pm$1 & $-0.03\pm0.10$ & 834 & May 6 2000 & 34 \\
& & & & & & 9569 & Feb 12 2008 & 101 \\
NGC\thin 4472 & 15.1 & 1.7  & 294$\pm$3& 9$\pm$2 & $0.17\pm0.12$ & 321 & Jun 12 2000 & 34 \\
\enddata
\tablecomments{Basic data for the objects in our sample. Distances were 
derived from the SBF survey of \citet{tonry01}, correcting the distance modulus 
by -0.16 mag to account for revisions to the Cepheid zero-point \citep[\eg][]{jensen03}. Total 
J-band luminosities (\lj) were derived from our own modelling (\S~\ref{sect_stars}). The 
central velocity dispersion of the stars (\sigmac) was taken from \leda\ and the 
age and mean metallicity ($<$[Z/H]$>$) were derived from our Lick index analysis
\citep[][\citetalias{humphrey06a}]{humphrey05a}.
The \chandra\ observation identification number
(ObsID) is given for each dataset, and we list the date the observation was made 
and the total exposure time following removal of background flare events.}
\end{deluxetable*}
To select our targets for this study, we searched the
\leda\ database for early-type galaxies (T$\le$-1) with heliocentric recessional
velocities ${\rm <2300 km\ s^{-1}}$ (corresponding to a distance $\sim$33~Mpc) and
central velocity dispersion, \sigmac\ ${\rm >280 km\ s^{-1}}$, \citep[corresponding to a black hole mass 
$\sim 5\times 10^8$\msun;][]{tremaine02a}.
Restricting ourselves to gas-rich galaxies that have been observed with \chandra,
our sample contained 9 objects. We omitted three galaxies 
(M\thin 87, NGC\thin 3998 and IC\thin 1459)
as they contain X-ray emitting AGNs which may make ISM temperature measurements at \ltsim 2\arcsec\
scales challenging. This left 6 objects: NGC\thin 1332, NGC\thin 1399, 
NGC\thin 4261, NGC\thin 4374, NGC\thin 4472 and NGC\thin 4649. NGC\thin 4261 also hosts an 
X-ray emitting AGN but it is highly absorbed, so that the thermal gas dominates the spectrum below
$\sim$2~keV, enabling accurate gas temperature determination even at the smallest scales
(\S~\ref{sect_spectra}; \citealt{zezas05a}; \citetalias{humphrey06a}).
{We excluded NGC\thin 4374 since it has a very disturbed
X-ray morphology, in comparison to the other systems
\citep{finoguenov01}, making accurate spectral deprojection very 
challenging.}
Although \citet{churazov08a} argued that the gas
in NGC\thin 1399 is relatively close to hydrostatic, we also excluded that system on account of its 
disturbed morphology at all scales \citep{scharf05a} and the presence of multi-phase gas
\citep{buote02a,humphrey05a}, both of which may
 introduce systematic uncertainties into our measurements.
We have already published the results for NGC\thin 4649 \citepalias{humphrey08a}; this paper 
focuses on the three remaining objects (although we also discuss 
updated results for NGC\thin 4649,
having applied the refined analysis methods outlined in the present work).
 We list the basic properties of these three galaxies in 
Table~\ref{table_sample}.

We have previously published an X-ray mass analysis for both NGC\thin 4261 and NGC\thin 4472
\citepalias{humphrey06a}, and found that the gas appeared close to hydrostatic. However, our
previous analysis focused primarily on the properties of the dark matter.
Moreover we adopted a simple, ad hoc form for the stellar mass profile, which 
introduced considerable systematic uncertainty into the stellar mass-to-light (M/L) ratio
measurement and clearly precludes the accurate determination of a central black hole mass.
In this paper, we therefore
present a new analysis for these galaxies, 
using an updated treatment of the stellar mass model, and the 
improved mass-determination methods outlined
in \citetalias{humphrey08a}. NGC\thin 1332 is an X-ray luminous
lenticular galaxy with a relaxed  X-ray morphology and a 
centrally peaked gas temperature profile \citep{buote96a,humphrey04b}.
Of the three galaxies, only NGC\thin 1332 does not have a published \mbh\
measurement derived from dynamical methods. Both NGC\thin 4261 and 
NGC\thin 4472 have \mbh$\sim 5\times 10^8$\msun, determined from, respectively,
gas kinematics \citep{ferrarese96a} and stellar dynamics \citep{gebhardt09a}.
\section{X-ray data analysis} \label{sect_analysis}
\subsection{Data reduction}
\begin{figure*}
\plotone{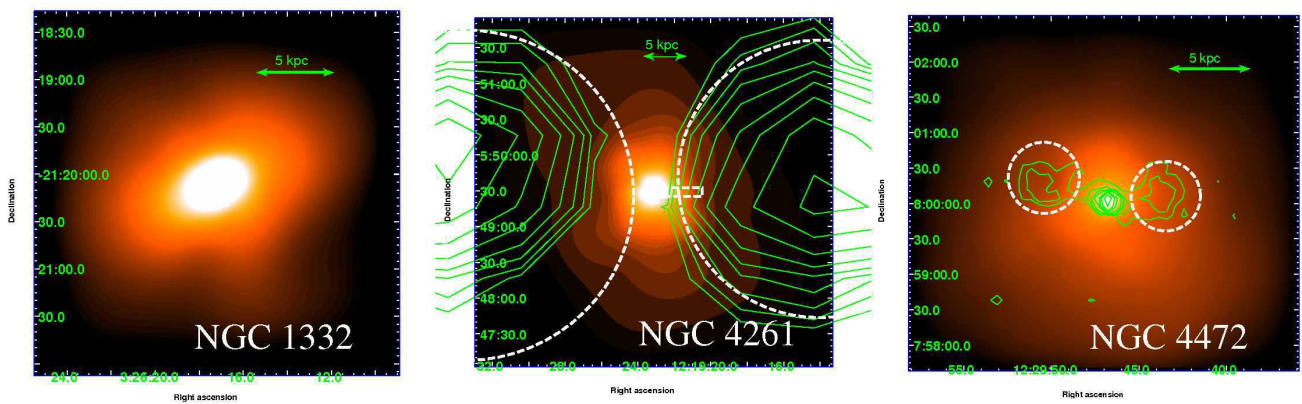}
\caption{Smoothed, point-source subtracted \chandra\ X-ray images of the galaxy sample
in the 0.5--7.0~keV band (see text). These images only cover the central part of the 
ACIS-S3 chip, where the count-rate is sufficiently high for the structure of the X-ray
emission to be clearly discerned.
All three galaxies show relatively 
relaxed morphologies, although there is evidence of cavities in NGC\thin 4472
and NGC\thin 4261. {The image
contrast and smoothing scales were arbitrarily adjusted to bring out key features;
the smoothing scale varies from $\sim$1\arcsec\ in the central part of the images to 
$\sim$1\arcmin\ in the outer regions.}
We overlay radio contours from the NVSS survey on the image of NGC\thin 4261
and VLA FIRST survey contours on the \chandra\ image of NGC\thin 4472. We also
mark regions of the field of view which were excluded from our analysis (shown circumscribed
by dotted white lines). 
\label{fig_images}}
\end{figure*}
The region of sky containing each galaxy was imaged by the \chandra\ ACIS instrument
in the ACIS-S configuration. The details of each observation are given in Table~\ref{table_sample}.
The archival data were reprocessed using the CIAO~4.0 (beta 3) and \heasoft\ 5.3.1
software suites, in conjunction with the \chandra\ calibration database (\caldb) version 3.5.0.
To ensure up-to-date calibration, all data were
reprocessed from the ``level 1'' events files, following the standard
\chandra\ data-reduction threads\footnote{{http://cxc.harvard.edu/ciao/threads/index.html}}.
We applied the standard correction to take account of the time-dependent gain-drift
and charge transfer inefficiency,
as implemented in the \ciao\ tools. To identify periods of enhanced
background (which can seriously degrade the signal-to-noise, S/N)
we accumulated background lightcurves for each dataset from
low surface-brightness regions of the active chips, excluding obvious
point-sources. Such periods of background ``flaring'' were 
identified by eye and
excised. The final exposure times are listed in Table~\ref{table_sample}.
{For NGC\thin 4261, we processed both datasets independently, and then merged the final
products (images, spectra, spectral response files, \etc) using the procedure outlined in
\citetalias{humphrey08a}, which involves correcting for relative astrometric errors
by comparing positions of the detected point-sources.}

For each
galaxy we generated a full resolution image in the 0.5--7.0~keV energy-band
and a corresponding
exposure map computed at an energy of 1.7~keV.
Point sources were detected with the \ciao\ {\tt wavdetect} task,
which was set to search for structure at scales of 1, 2, 4, 8 and 16
pixels, and supplied with the exposure-map to minimize spurious detections at the image boundaries.
The detection threshold was set to $10^{-6}$, corresponding to \ltsim 1 spurious source
detections per chip. All detected sources were confirmed by visual inspection, and, for each,
appropriate elliptical regions containing approximately 99\%\ of its photons were generated.
In Fig~\ref{fig_images} we show  smoothed, flat-fielded \chandra\ images of each galaxy, having removed 
the point-sources from the images. To produce these images we first replaced all the photons within the source 
detection ellipse of each source with artificial data, using the algorithm described in 
\citet{fang09a}.
The data were then flat-fielded with the 
exposure map and smoothed with a Gaussian kernel. Since the S/N of the image varies strongly
with distance from the centre of the galaxy, the width of the Gaussian kernel was varied
with distance according to an arbitrary power law. For each galaxy, the X-ray image shows
a relatively relaxed, centrally peaked morphology within the central 
$\sim$10~kpc (where the stars are expected to be the dominant mass component).

In Fig~\ref{fig_images} we overlay the radio contours from  the NVSS survey \citep{condon98a} 
on the NGC\thin 4261 image and those from the VLA FIRST survey \citep{becker95a} on the image of NGC\thin 4472. 
In both galaxies the radio lobes appear coincident
with slight depressions in the X-ray brightness, which may be ``cavities'' 
\citep{biller04a,croston05a}. Although we have previously shown \citepalias{humphrey06a} that the gas
density and temperature can be fitted with physically meaningful hydrostatic solutions even
if data from the vicinity of these cavities is included, to minimize potential 
systematic uncertainties in our subsequent analysis we opted to mask them out,
as shown in Fig~\ref{fig_images}. We also masked out data around the jet in NGC\thin 4261
\citep{zezas05a}.

\subsection{Spectral analysis} \label{sect_spectra}
\begin{figure*}
\plotone{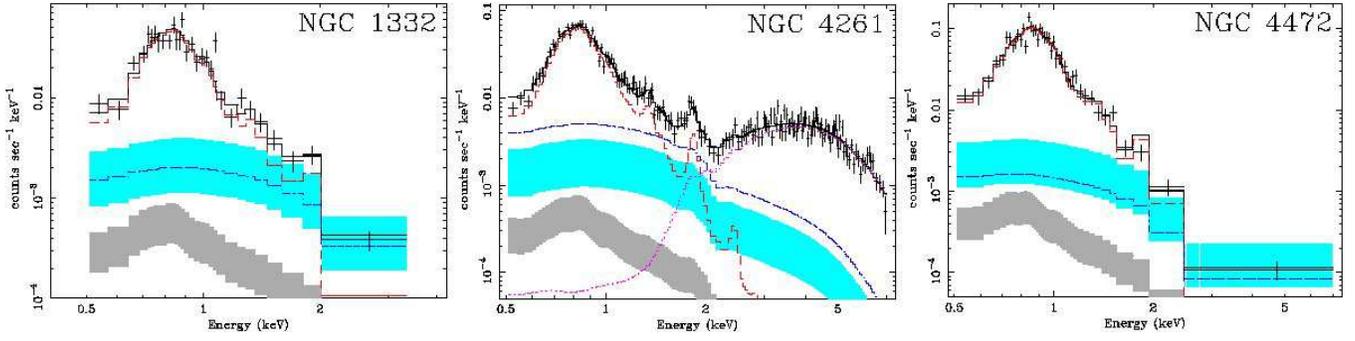}
\caption{X-ray spectrum measured in the central bin of each galaxy. We show the 
measured data (without removing instrumental features), and the best-fitting 
model, folded through the instrumental response (solid black line). We also show
individually the modelled contributions of the hot gas (dashed red line), the unresolved
LMXBs (dash-dot blue line) and, for NGC\thin 4261, the central AGN
(dotted magenta line). Given known correlations between the K-band light and the 
luminosity of unresolved point sources, we show the region in which the LMXB
emission is expected to lie at 1-$\sigma$ uncertainty (upper, light blue shaded region),
and the region in which the contribution
from the composite X-ray emission of cataclysmic variables and stars is expected to
lie (lower, grey,
shaded region). Hot gas unambiguously dominates the spectra below $\sim$2~keV in all three galaxies.
\label{fig_spectra}}
\end{figure*}
We extracted spectra in a series of concentric, contiguous annuli, placed at the X-ray centroid
of each object.  We determined
the centroid iteratively by placing a 0.5\arcmin\ radius aperture at the nominal
galaxy position (obtained from \ned) and computing the X-ray centroid
within it. The aperture was moved to the newly computed centroid, and the
procedure repeated until the computed position converged. The final centroids were all within
$\sim$0.6\arcsec\ of the optical centre, estimated from our HST data (\S~\ref{sect_stars}),
having corrected for differences in the absolute astrometry by matching X-ray and optical 
point sources \citep{humphrey08b}.

The widths of the annuli were chosen so as to contain approximately the same number of 
background-subtracted photons in each (we relaxed this criterion for the innermost bins 
to provide as high
spatial resolution as possible), and to ensure there were sufficient photons to perform useful
spectral-fitting. 
We placed a lower limit of 2.5\arcsec\ on the annulus width (for the circular
central bin, this refers to its {\em diameter}), to ensure
the finite instrumental spatial resolution does not lead to strong mixing between the spectra in 
adjacent annuli. The data in the vicinity of any
detected point source were excluded, as were the data from the vicinity
of chip gaps, where the instrumental response may be uncertain. 
We extracted
products from all active chips, excluding S4 (which suffers from
considerable ``streaking'' noise). 
Appropriate count-weighted spectral response
matrices were generated for each annulus
using the standard \ciao\ tasks {\tt mkwarf} and {\tt mkacisrmf}. For each spectrum,
we estimated the background using the method outlined in \citetalias{humphrey06a}.

\begin{figure*}
\plotone{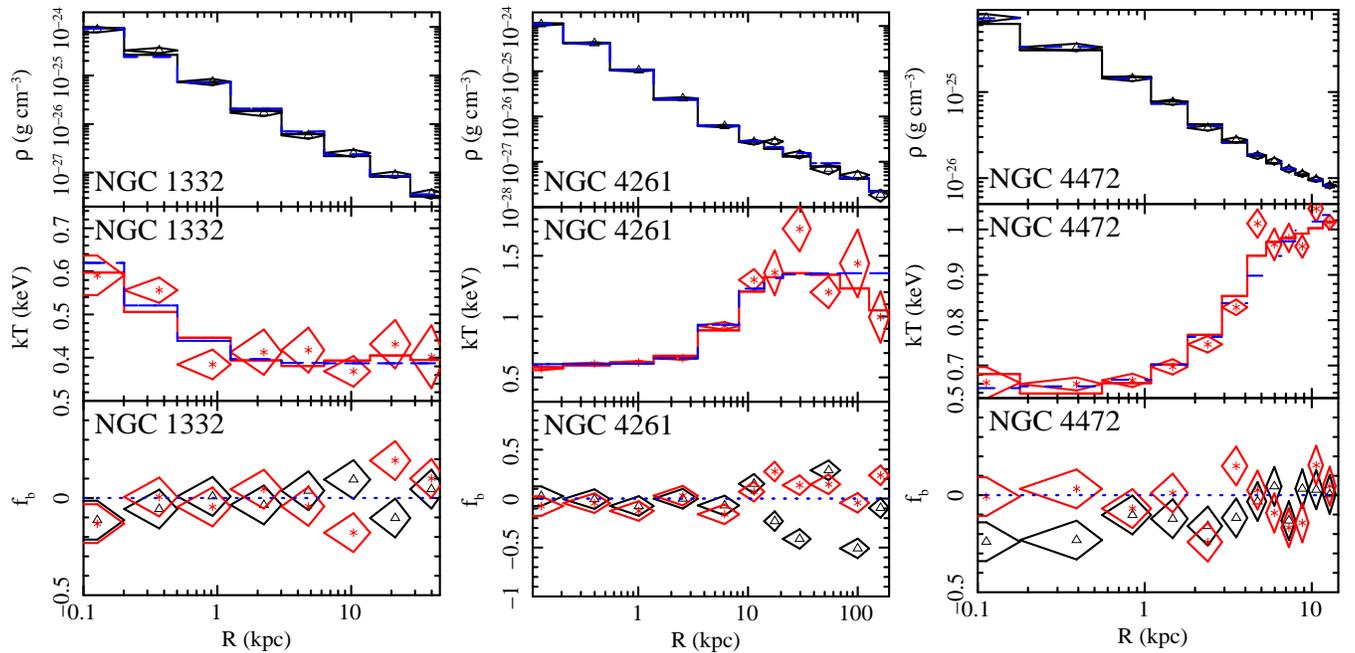}
\caption{Density ($\rho$; triangles) and 
temperature (kT; asterisks) profiles for the hot gas in each galaxy.
We also show the bias fraction \fb\ (see text) for both kT and $\rho$, which indicates
that the derived data are largely unbiased. The best-fitting 
density and temperature models described in \S~\ref{sect_results} are shown as 
solid lines and fit the data well. We also show (dotted lines) the best-fitting arbitrary
parameterized models for the density and temperature used in our ``traditional''
mass analysis (\S~\ref{sect_nonparametric}). {All error-bars correspond to 1-$\sigma$
uncertainties.}
\label{fig_profiles}}
\end{figure*}
Spectral-fitting was carried out in the energy-band 0.5--7.0~keV, using the 
 XSPEC spectral-fitting package \citep{xspec}. We have shown previously
 that fits to Poisson distributed data which minimize $\chi^2$ can yield 
significantly biased results,
even if we rebin the data to more than the canonical $\sim$20 counts 
per bin \citep{humphrey09b}.
In contrast, we found that the C-statistic of \citet{cash79a} typically
gives relatively unbiased results.
We therefore performed the fits by minimizing C, but nonetheless 
took care to estimate any residual bias on
our fits by using the Monte Carlo technique outlined in \citet{humphrey09b}, as discussed in 
detail below. Although not strictly necessary for a fit using the C-statistic, we rebinned
the data to ensure a S/N ratio of at least 3, and a minimum of 20 photons per bin. 
Such rebinning is 
useful during fitting as it emphasizes differences between the data and the model if
 the fit is poor.

For each galaxy, we fitted the data from all annuli simultaneously. This allowed  us to take 
advantage of the {\bf projct} model in \xspec, which enables spherical deprojection. 
Spherical symmetry is typically assumed in X-ray mass analysis, even if the X-ray isophotes are not perfectly 
round, as deviations from this approximation are
 only expected to contribute a relatively small error into the
recovered mass profile \citep[\eg][]{piffaretti03a,gavazzi05a}. We will return to the possible impact of asphericity
in a future paper; for now, however, we are interested in how accurately a conventional
spherical, hydrostatic mass analysis can recover the mass profile in the inner part of the galaxy.
To take account of projected emission from regions outside the outermost annulus we used the 
procedure outlined in \citetalias{humphrey06a}.
To model the hot gas we adopted a {\bf vapec} component, plus a 7.3~keV
thermal bremsstrahlung component to account for  undetected low-mass X-ray binaries (LMXBs)
\citep[this model gives a good fit to the composite spectrum of the detected LMXBs in nearby
galaxies:][]{irwin03a,humphrey08b}.  For NGC\thin 4261, we included an additional absorbed powerlaw
component in the central bin to account for the AGN emission \citepalias{humphrey06a}.
We adopted a slightly modified form of the \xspec\ {\tt vapec} code to enable us to 
tie the ratio of each elemental abundance with respect to Fe between each annulus, 
while the Fe abundance (\zfe) was allowed to fit freely. 
To improve S/N, \zfe\ was tied
between adjacent annuli where necessary.
We allowed the global ratios of O, Mg and Si with respect to Fe to fit freely, while the
remaining ratios were held at their Solar values \citep{asplund04a}. For NGC\thin 4472, we 
additionally freed S and Ni and for NGC\thin 1332 and NGC\thin 4261 we freed Ne. 
The absorbing column density (\nh) 
was fixed at the Galactic value \citep{dickey90}.
The best-fitting abundances 
were consistent with those reported  
in \citet{humphrey05a} and \citetalias{humphrey06a}.

The X-ray spectrum of the innermost bin for each galaxy is shown in Fig~\ref{fig_spectra}, along with 
the best-fitting model, folded through the instrumental responses, and the contribution of each of the 
spectral components. Overlaid we show the {\em expected} spectrum of the unresolved
LMXBs (upper shaded region), and the expected integrated X-ray emission of cataclysmic variables and stars (lower shaded region). These were estimated
from the relevant correlations with the K-band light 
reported by, respectively, \citet{humphrey08b} and \citet{revnivtsev08a}\footnote{Following
\citet{revnivtsev08a}, for the spectrum of the CVs and stars 
we use a {\bf MEKAL} plasma with kT=0.5, Solar abundances relative to \citet{anders89}, plus a 
$\Gamma=1.9$ powerlaw component. We fixed  the ratio of the total flux in the 2.0--7.0~keV band to the that in 
the 0.5--2.0~keV band to be 0.77.}.
Since neither correlation with the K-band light is exact, we shade the region allowed by the 1-$\sigma$
scatter in the relations. As is clear from the figure, the measured 
normalization of the bremsstrahlung component 
is consistent (within $\sim$1.8--$\sigma$) with the LMXB flux 
predicted from the relation of \citet{humphrey08b}
for all three systems.
The K-band magnitude was actually estimated from the I-band HST data (\S~\ref{sect_stars}), and corrected
for the difference in filters by 
using the difference in magnitudes measured between the I-band data and K-band \twomass\
data in a $\sim$7\arcsec\ aperture. {\em Clearly in all three galaxies, the 
X-ray emission in the innermost bin is completely  dominated (below $\sim$2~keV) by the hot gas emission}.
This is also true for the other annuli.

Error-bars were computed {\em via} 
the Monte-Carlo technique outlined in \citetalias{humphrey06a}, and we carried out 100 error simulations. 
We show the measured kT and gas density, 
\rhog\ (which can be computed from the normalization of the {\bf vapec} component) as a function of 
radius in Fig~\ref{fig_profiles}. 
As discussed in \citet{humphrey09b}, such simulations can be used to assess whether there is any bias
on the best-fitting results by evaluating \fb\ (the ratio of the bias to the statistic
error) for each parameter. 
This is computed by taking the difference
of the best-fitting value of each parameter and the mean value of that parameter
obtained from the simulations, and then 
dividing this result by the statistical error. If $|$\fb$|\ll 1$ the fits are 
practically unbiased. In Fig~\ref{fig_profiles} we show \fb\ for kT and \rhog\ in
each annulus. We found that, in general,
$|$\fb$|\ll 1$, supporting our use of the C-statistic. The temperature and density appeared 
slightly biased only in to two annuli in NGC\thin 4261, for which the error-bars are large and hence
these data are unlikely to drive the temperature and density fits discussed in \S~\ref{sect_models}. 
The origin of the bias is unclear, but we suspect it may be related to ``deprojection noise''. 

\section{The stellar mass distribution} \label{sect_stars}
\begin{figure*}
\plotone{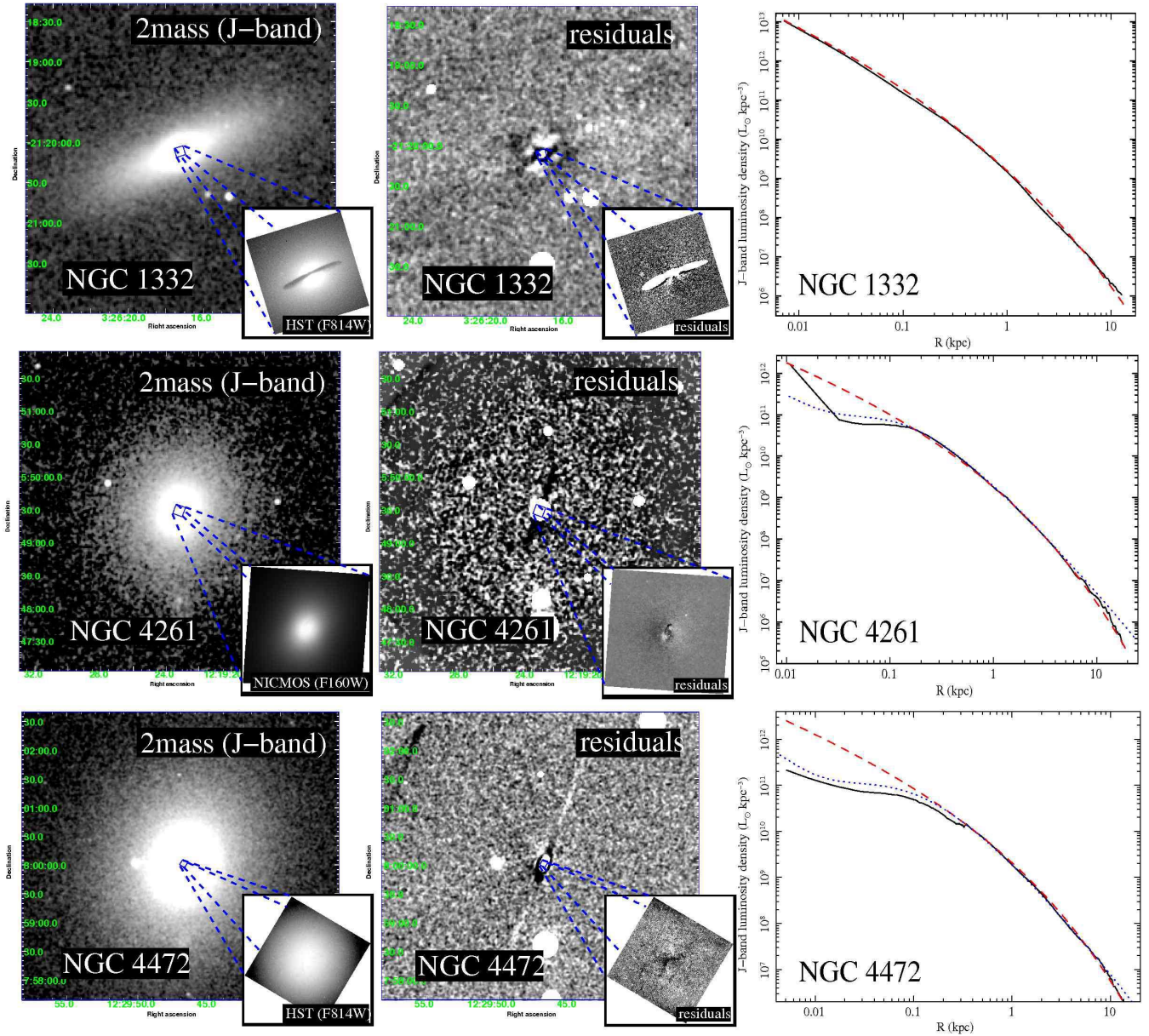}
\caption{Left panels: J-band \twomass, I-band \hst\ WFPC2 and H-band NICMOS 
images of each galaxy. For the HST data we annotate the image with the appropriate HST filter.
Central panels: residual images, having subtracted our projected 
model for the three-dimensional light distribution (see text). Regions which were excluded
from the deprojection are shown as white ellipses. Overall, these models provided a good 
fit to the optical light, {except for the central part of NGC\thin 4261, where residual dust
extinction is problematical. Right panels: the spherically averaged luminosity density as a function
of geometric radius from our model (solid lines). For NGC\thin 4261 and NGC\thin 4472,
we overlay (dotted lines) deprojected V-band profiles light profiles, arbitrarily 
scaled, obtained by applying the algorithm of \citet{gebhardt96a} to the 
data of \citet{kormendy09a}. These profiles agree well with those obtained using our 
default deprojection method.}
We also show the best-fitting deprojected
Sersic model parameterization (dashed lines) for these profiles (see text). 
\label{fig_optical_images}}
\end{figure*}
The modelling technique we have introduced for analysing the X-ray data of early-type galaxies
\citepalias{humphrey08a} relies on an accurate (3-dimensional) representation of the stellar mass component.
This is particularly true when one wishes to measure the mass of the central black hole, since one requires
very accurate subtraction of the stellar mass. To determine the stellar
light density, we adopted an axisymmetric deprojection technique; this is 
particularly important given the highly elliptical isophotes of NGC\thin 1332,
which is an almost edge-on lenticular galaxy. Although
we subsequently spherically averaged the three-dimensional stellar density
for use in our mass modelling (\S~\ref{sect_models}), we 
adopted this procedure to ensure the spherically averaged profile
was as accurate as possible.  
Specifically we used
the iterative axisymmetric deprojection algorithm
pioneered by \citet{binney90a}. {This is a very general technique for
axisymmetric deprojection, which is well-suited for modelling
isophotal shapes ranging from highly elliptical (as in NGC\thin 1332)
to very boxy (as in the central part of NGC\thin 4261), as well
as systems in which the projected isophotal 
ellipicity varies strongly as a function of radius,
as is generally the case here \citep{kormendy09a,buote96a}.
As an axisymmetric method it cannot, however, account for the mild
isophotal twists which are seen, so there will be some 
associated uncertainty in the deprojected light profile.}
Briefly, this technique involves
first fitting an arbitrary function to the images, which is 
analytically deprojected onto an (r,$\theta$) grid
as a ``first guess'' for the true stellar light distribution 
(here the light is assumed to be axisymmetrically distributed, with coordinate 
r indicating the distance from the centre of the galaxy and $\theta$ 
the zenith angle).  A series of Lucy-Richardson ``relaxation'' 
iterations \citep{lucy74a} subsequently improves the estimate nonparametrically.

{To obtain the optical light profile, we used \twomass\ J-band images and, to provide superior spatial resolution
in the crucial inner parts of the images, 
\hst\ WFPC2 data using the F814W filter (approximately I-band) for NGC\thin 1332
and NGC\thin 4472, and NICMOS2 data, using the F160W filter (approximately H-band) 
for NGC\thin 4261. The NICMOS data were used in preference to WFPC2 data 
for NGC\thin 4261 to minimize attenuation due to
the central dust disk, which complicates the deprojection.
Since we aimed to model the temperature and data density points measured with \chandra,
for our purposes we required only the total stellar light enclosed within a region no
smaller than $\sim$1\arcsec\ (\S~\ref{sect_spectra}) and it was therefore not 
necessary to deconvolve the instrumental PSF from the HST data.
The preparation of the WFPC2 data is described
in \citet{humphrey09a}, although for NGC\thin 4472 we used different
data than those listed therein
(specifically we used observation numbers u2lg0603t and u2lg0604t, for which the galaxy is better centred on the 
PC chip).  For the NICMOS data, we used the publicly available,
pre-cleaned image described in \citet{floyd08a}. To prevent the emission from
the weak AGN that is visible in the NICMOS image of NGC\thin 4261 from introducing bias,
we replaced the data within a central $\sim$0.35\arcsec\ region with the mean value of the 
surrounding pixels \citep[which is reasonable given the flat surface brightness
profile in this part of the galaxy:][]{kormendy09a}. 
In practice we did not use the entire WFPC2 and NICMOS images, but simply a 
portion close to the centre of the galaxy, 
chosen to overlap the area where seeing seriously
 compromises the \twomass\ data.}
The images are shown in Fig~\ref{fig_optical_images}.

To obtain the initial guess for the stellar density, we simultaneously
fitted models to the \hst\ and \twomass\ images 
using custom-made software built around the \minuit\
library\footnote{http://lcgapp.cern.ch/project/cls/work-packages/mathlibs/minuit/index.html}. We masked out
obvious point-sources and the central parts of the \twomass\ images, where 
the seeing is problematic. 
{The WFPC2 images of NGC\thin 1332 and NGC\thin 4261 revealed 
central dusty discs; for NGC\thin 1332, we masked this out as best 
we could.}
For NGC\thin 4472 and NGC\thin 4261 we were able to fit the data 
satisfactorily with a model which is the analytical projection of a 
spherically symmetric Sersic density profile \citep{prugniel97a}, having
broken the symmetry by arbitrarily scaling one of the coordinate axes;
\ie\ $\rho_*(x,y,z) = a \rho_{*,s} (\sqrt{x^2+(ay)^2+z^2})$. Here
$\rho_*(x,y,z)$ is the luminosity density profile, $a$
is an arbitrary scaling parameter and $\rho_{*,s}$ is the deprojected
Sersic density profile. In projection, this produces elliptical isophotes, 
the ellipticity
of which depend on both the factor $a$ and the inclination angle $i$. In
general $i$ is not known, and so we arbitrarily adopted $i=90^\circ$.
We discuss the impact of changing the value of $i$ in 
\S~\ref{sect_systematics}. We allowed the Sersic index, effective radius, 
scaling factor $a$
and position angle of the isophotes to fit freely, but constrained them
to be the same for both the \twomass\ and \hst\ data of a given galaxy,
taking into account orientation differences between the images.
The centroids of the models were allowed to fit freely in each image, to
allow for possible errors in the absolute astrometry.
We found there were two best-fit solutions which could not be distinguished
statistically--- one with 
$a>1$ (oblate) and one with $a<1$ (prolate). By default we adopted the oblate
model, but consider the effect of using the prolate model in 
\S~\ref{sect_systematics}. 
For NGC\thin 1332, the data required two such Sersic models,
plus an inclined exponential disc (which can be deprojected trivially,
under the assumption of an arbitrarily small but finite opening angle for 
the disc). We required the position angle of the Sersic models and the 
disc to be the same, and similarly tied the inclination of each component.
 Since the ellipticity of the disc model depends only on $i$, 
this model allowed us to constrain $i$ to $\sim73^\circ$.

Having obtained the initial fits to the surface brightness profile, and hence
an initial guess for the deprojected density, we used the relaxation
algorithm of \citet{binney90a} to refine our estimates. Although the algorithm
they discussed is relevant for a single image, it is straightforward to generalize
it to a stack of images with regions masked out. Since an image is simply
a list of pixels, one always has the freedom to define a new ``image'',
every pixel of which maps uniquely to one of the pixels in the image stack.
Not every pixel in the image stack need have a counterpart in the new 
``image'', so that one can readily mask out problem regions.
Although this transformation complicates the definition of the projection 
kernel, in principle exactly the same methods can be applied to 
deproject this new ``image''. 
In practice, we also subtracted the fitted background level from the count-rate in
each pixel before mapping it onto the new ``image'', 
resetting to zero any pixels which fall below 0 counts, to prevent
erroneously treating the background as a part of the source emission.

After each relaxation iteration, the ``improved'' density profile was 
projected onto the \twomass\ and \hst\ image planes. We fitted this new
``model'' to the data, allowing the overall normalization, the scaling
 factor between the \hst\ and \twomass\ data and the background 
levels to fit freely. We continued iterating in this manner until the 
reduction in the fit statistic on any given iteration was less than 1,
at which point we assumed convergence.
In Fig~\ref{fig_optical_images} (middle
column) we show the residuals (data minus model) obtained from our 
deprojection technique for each galaxy. {In general, the residuals are small,
except for the central arcsec of NGC\thin 4261, where the dust disc is visible
even in the NICMOS data. Due to the residual dust extinction, it seems likely that our 
deprojected profile for this galaxy slightly underestimates the true central stellar density;
we investigate the possible impact of this underestimate in
\S~\ref{sect_systematics_stars}}. In the right hand column of 
Fig~\ref{fig_optical_images} we show the spherically averaged density profile 
for each system, defined as the mean luminosity density within a thin,
spherical shell of given radius. 
{We also show (dashed lines) 
the best-fitting  spherical Sersic approximation to our deprojected profiles
(the parameters of these fits are shown in the figure). These are shown
to guide the eye since, by default, we use the Lucy deprojected profiles 
in our analysis (although in \S~\ref{sect_systematics_stars} 
we investigate how significantly our results
are changed if we use the Sersic profiles shown in Fig~\ref{fig_optical_images}).
For these fits, we find \reff=2.3~kpc, 3.0~kpc and 4.2~kpc, respectively for
NGC\thin 1332, NGC\thin 4261 and NGC\thin 4472, while the Sersic index, n, is 
4.2, 3.1 and 2.9, respectively. The values of the Sersic index we obtained
for NGC\thin 4261 and NGC\thin 4472 are lower than found by \citet[][]{kormendy09a}, 
who reported values of 7.5 and 6.0, respectively, based on V-band surface brightness fits.
The reason for the discrepancy is that we fitted the entire inner profile including the core,
 but did not extend the fit to large radii, while \citeauthor{kormendy09a} fitted out to very large 
radii, but excluded the core region, which they considered to be an interesting departure from the model.}

{To test further the sensitivity of our results to the deprojection method,
for NGC\thin 4261 and NGC\thin 4472 we also
used the  algorithm of \citet{gebhardt96a} to deproject the 
the V-band surface brightness profiles in \citet{kormendy09a}. We assumed that the
stellar isodensity surfaces can be represented by similar, axisymmetric spheroids
(with axis ratios 0.75 and 0.81 for NGC\thin 4261 and NGC\thin 4472, respectively)
and adopted an inclination angle of 90$^\circ$. The resulting, spherically averaged
density profiles are shown (arbitrarily scaled, for clarity) 
in the right column of Fig~\ref{fig_optical_images} as dotted (blue) lines. Clearly there is 
good agreement with our best-fitting profiles over a wide radial range. The V-band 
profiles have a higher density outside $\sim$10~kpc, which may arise from imperfect 
sky subtraction in the \twomass\ data. Since the stellar mass density only enters into
our calculations through the total enclosed gravitating mass profile, which is dominated by the 
dark matter halo outside a few kpc \citepalias{humphrey06a}, this does not pose a 
significant problem for our analysis.
We discuss the effect of using the deprojected \citeauthor{kormendy09a} profiles
on our results in \S~\ref{sect_systematics_stars}.}


\section{Mass modelling} \label{sect_models}
The temperature and density profiles of gas in hydrostatic equilibrium are uniquely determined by 
four factors--- the mass profile, the profile of the entropy proxy
($s=kT n_e^{-2/3}$, where $n_e$ is the electron density)
the gas pressure at a fiducial radius {and the total gas mass enclosed within
the fiducial radius. We choose sufficiently small a fiducial radius that we
can assume the enclosed gas mass to be zero.}
One can, therefore, invert the problem and use the temperature and density profiles to 
recover the mass. We have found the most robust way to 
achieve this is a ``forward fitting'' method, where physically motivated parameterized models for 
the mass distribution and the entropy profile are used to predict the density and temperature profiles,
which are then directly fitted to the data (\citetalias{humphrey08a}; 
also see \citealt{gastaldello07a} for some 
variations on this approach).

Assuming spherical symmetry,
we modelled the enclosed mass profile as a stellar component (which was assumed to be proportional to the 
J-band luminosity enclosed in a sphere of given radius)
plus an NFW \citep{navarro97} dark matter component, a 
central supermassive black hole,
and the self-consistently calculated gas mass.
We allowed the stellar mass-to-light (M/\lj)
ratio to be fitted freely, along with the black hole mass (\mbh) and total mass and concentration
of the dark matter halo. 
We modelled the entropy as a broken powerlaw plus a constant, \ie:
\begin{equation}
s  =  s_0 + s_1 f(r) \label{eqn_entropy}
\end{equation}
where:
\begin{eqnarray}
f(r)  = &  \left( \frac{r}{10~kpc} \right)^{\beta_1}  & (r < s_{brk}) \nonumber \\
 & \left( \frac{s_{brk}}{10~kpc} \right)^{\beta_1-\beta_2}  \left( \frac{r}{10~kpc} \right)^{\beta_2}  & (r \ge s_{brk})
\end{eqnarray}
and the parameters $s_0$, $s_1$, $\beta_1$, $s_{brk}$ and $\beta_2$ were 
parameters of the fit. Furthermore, we allowed the 
central gas pressure to be determined by our fit. For a more detailed description of the modelling
procedure, see \citetalias{humphrey08a}. 

As is standard practice, we assumed that each temperature and density data-point derived from our 
spectral-fits was a Gaussian-distributed, independent random variable.
This enabled us to compare the models to the temperature and density profiles by computing the $\chi^2$
fit statistic \citep[since the data were not Poisson distributed, we do not
expect the $\chi^2$ fits to be biased as described 
in][]{humphrey09b}.
In \citetalias{humphrey08a} we computed the entropy profile 
from the temperature and density data-points, and fitted the models
to the entropy and temperature profiles simultaneously. In the present 
work, however, we preferred to fit the models to the density and temperature
as they are the quantities derived directly from the spectral fits.
Clearly this is functionally the same thing; if a model fits the temperature
and density profiles well, it will also fit the entropy.
For NGC\thin 4472, we only considered data within $\sim 3.2$\arcmin ($\sim$14~kpc) as there
is evidence of X-ray asymmetries outside this range \citep{irwin96a,biller04a}. Although we have previously
shown that reasonable hydrostatic solutions can be constructed even when including the \chandra\ data at larger 
radii \citepalias{humphrey06a}, we did not include those data here so as to minimize any potential 
source of systematic uncertainty. To obtain robust measurements of the fit parameters, rather than simply 
minimizing $\chi^2$  we adopted a Bayesian  approach, as discussed below.

\subsection{The Bayesian prior} \label{sect_prior}
As with all Bayesian analysis, the choice of prior is a matter of considerable import. While for some of the
fit parameters (below) this choice is natural, for  other parameters
(the central gas pressure and the
parameters characterizing the entropy profile), it is less clear. In those cases it is tempting to resort 
to a flat prior, but this can be problematical. Since a prior which is defined to be flat for a given parameter will
no longer be flat under arbitrary transformation of that parameter (for example, taking the logarithm), 
even a ``flat'' prior adopted in ignorance arbitrarily imposes our preconceived notions on the fit.
Fortunately, this choice does not always have a significant impact on the derived results; in some cases, the data may actually place much tighter 
constraints on the parameter than any reasonable prior, and so a flat prior is as good
a default choice as any.
(Formally this is the case if the dynamical range of the parameter allowed by the $\chi^2$ 
likelihood function is sufficiently narrow that the Jacobean introduced by a reasonable parameter transformation
is approximately constant). Furthermore, if one is uninterested in the particular value of a given
parameter (\ie\ intending to marginalize over it), the choice of prior is also relatively unimportant 
{\em provided that parameter has little or no covariance with any parameter of interest}.
Nevertheless, it is unlikely that one will know that the problem satisfies these criteria before actually
exploring the Bayesian posterior, and so a reasonable approach is to cycle through multiple priors to examine the impact 
of each choice on the results. Adhering to this principle, by default we adopted flat priors on these problematical
parameters (\ie\ those describing the entropy profile and the pressure at the fiducial radius), and subsequently investigated
the effect of adopting alternatives (\S~\ref{sect_systematics_priors}).

For the other parameters the choice of prior is much clearer. A priori, we would expect that \mbh\ for each galaxy would
agree with the \mbh-\sigmac\ relation \citep{tremaine02a} and so we assumed a
$log_{10}M_{bh}$ prior which is Gaussian distributed about this relation, {given the
intrinsic scatter in this relation (taken to be 0.25~dex)
and the error on \sigmac\ (Table~\ref{table_sample}). We discuss the sensitivity of our
results to the form of this relation in \S~\ref{sect_systematics_priors}.}
Additionally, we have also explored the 
effect of using a flat prior for \mbh\ which, although poorly motivated, 
is useful for disentangling to what extent the measured
constraints are due to the prior and due to the data (\S~\ref{sect_results})\footnote{We note that another possible choice 
here, that of a prior which is flat on log \mbh\ (which is the same as a Prob(\mbh)=1/\mbh\ in linear space), 
would be a poor choice since it would imply {\em a priori}
that \mbh\ is as likely to be found in, for example, the range $10^3$--$10^4$\msun\ as the range $10^8$--$10^9$\msun.
Since the data for two of the galaxies have little discriminating power at very low masses (Fig~\ref{fig_prob_density}),
the use of this prior would unrealistically skew the black hole masses low and, crucially, will be sensitive to the 
lower mass cutoff. Conversely, a prior which is flat in \mbh\ produces a posterior probability distribution that is 
closer to that produced with the (more realistic) \mbh-\sigmac\ prior. Since the data tightly constrains the upper limit on
\mbh, the results using the flat prior are relatively insensitive to the upper mass cutoff.}.

Given the measured age and mean abundance of the 
stellar population in each galaxy (Table~\ref{table_sample}), 
stellar population synthesis models can be used to predict the 
stellar M/\lj\ and its uncertainty (\S~\ref{sect_mass_to_light}).
Although this  provides a natural prior for M/\lj,
it is of interest to compare the population synthesis predictions to those obtained from the fits.
Therefore, to avoid circular reasoning, we instead used a flat prior for M/\lj, 
but confirmed that consistent results were 
obtained with the more physical choice (\S~\ref{sect_systematics_priors}).
In practice M/\lj\ is sufficiently well constrained 
by the data that this choice has little
effect on the results.

If the galaxies were chosen entirely
at random (irrespective of morphological type or central velocity dispersion), one might simply adopt the halo mass
function predicted by dark matter simulations as the prior on the virial mass of the 
dark matter halo.
However this is clearly not appropriate here; a better choice would be to 
restrict the allowed range of the total baryon fraction 
within the virial radius, based on the predictions of a theoretical model,
as was done in \citetalias{humphrey06a}. Unfortunately, the physics of 
feedback is highly uncertain, making the results then sensitive to the assumptions
implicit in the adopted model. 
Alternatively, one could employ an empirical relation linking the 
mass of the galaxy to its host halo \citep[\eg][]{conroy08a}. We investigate this
option in \S~\ref{sect_systematics} but, in the interest of more generality in 
our results, by default  we used a uniform prior for the logarithm of the halo mass in the 
range $log_{10} M_{dm}=10^{12}$--$10^{14}$\msun.
For a prior on the concentration, we adopted the 
theoretical concentration--halo mass relation for relaxed halos found by 
\citet{maccio08a}. We used their relation for the ``WMAP1'' \lcdm\ cosmology since,
of the cosmologies they considered, that is most consistent
with observations of galaxies, groups and clusters
\citep{buote07a}.

\subsection{Exploration of the Bayesian posterior} \label{sect_multinest}
The exploration of the Bayesian posterior for a multi-parameter
model is typically carried out with Monte Carlo methods; we 
employed the publicly 
available\footnote{{http://www.mrao.cam.ac.uk/software/multinest/}} MultiNest code
{\em vers.}\ 1.0 \citep{feroz08b,feroz08a}, which implements a robust and
efficient nested sampling algorithm
\citep[a Monte Carlo integration
technique optimized for these purposes:][]{skilling04a}.
We have verified that we obtained consistent results with a Metropolis algorithm
based Markov Chain Monte Carlo code.
Since nested sampling integrates over a finite volume, in the absence of
hard boundaries for any parameters (such as the restricted range of dark matter
virial masses we considered)
we restricted each parameter to a $\pm 10$--$\sigma$ region about the 
``best-fitting'' value. This was estimated by first
minimizing $\chi^2$ with the Levenberg-Marquardt algorithm
\citep[\eg][]{nr}, which provides for each parameter 
both the ``best-fitting'' value and  an estimate of the statistical error, $\sigma$.
On completion of the nested sampling run, we checked for self-consistency between the initial estimates of $\sigma$ and the final
error-bars. If the latter were significantly larger than the initial guess, 
we re-ran the nested sampling algorithm with our refined
error estimate, thus ensuring at least a $\sim\pm$10--$\sigma$ region of integration for each parameter.

On completion, the nested sampling algorithm returns a set of data-points which sample the posterior reasonably well.
One can use these data to construct a probability density histogram for each parameter, marginalizing over the other parameters.
From this histogram, we found both the most probable values and a 
90\%\ confidence region, defined by stepping out
along lines of constant probability density until 90\%\ of the total probability was enclosed. We found that the results do not
depend sensitively on the width of the histogram used. 

\begin{figure}
\includegraphics[width=3.2in]{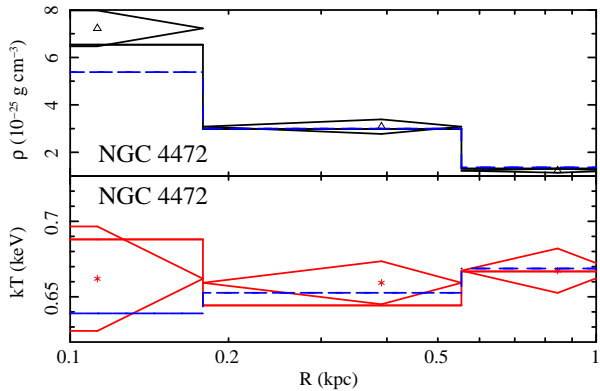}
\caption{Central part of the temperature and density profiles for the gas in 
NGC\thin 4472, illustrating the effect of adding a black hole. We show the 
best-fitting model with a non-zero \mbh\ (solid lines) and the best-fitting
model with \mbh=0. {The latter model is a poorer fit to both profiles in 
the central bin; the difference in $\chi^2$ between the two models is 5.5.
Error-bars correspond to 1-$\sigma$ uncertainties.}
\label{fig_n4472_profile}}
\end{figure}
\begin{figure*}[!f]
\plotone{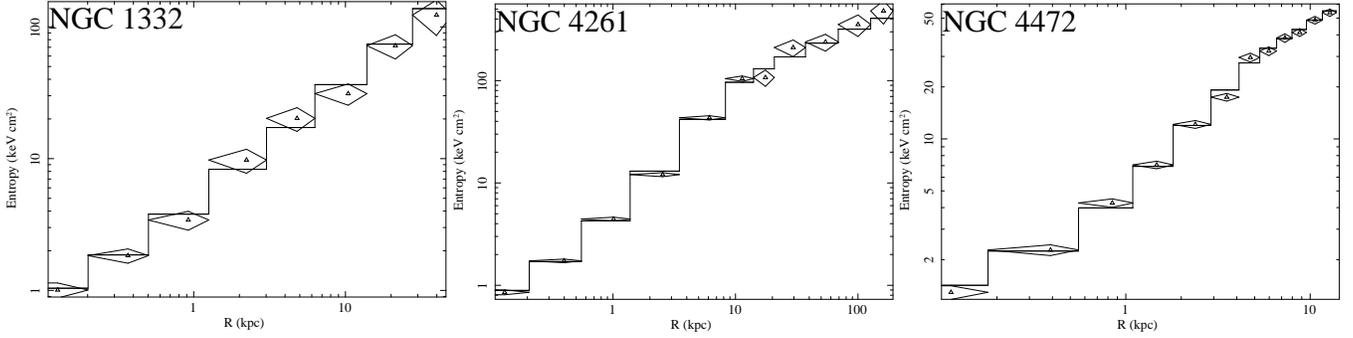}
\caption{Entropy profiles for each galaxy, derived from the density and 
temperature data in Fig~\ref{fig_profiles}. We also show the best-fitting
entropy models (\S~\ref{sect_results}), which fit the data very well.
{Error-bars correspond to 1-$\sigma$ uncertainties.}
\label{fig_entropy_profiles}}
\end{figure*}
\section{Results} \label{sect_results}
\begin{deluxetable*}{llrrrrrrr}
\tabletypesize{\scriptsize}
\tablecaption{Fit results \label{table_results}}
\tablehead{\colhead{Galaxy} & \colhead{$\chi^2$/dof} & \colhead{\mbh} & \colhead{M/L${\rm _*}$} & 
  \colhead{$s_0$} & \colhead{$s_1$} & \colhead{$\beta_1$} & \colhead{$s_{brk}$} &  \colhead{$\beta_2$} \\
\colhead{} & & \colhead{($10^9$\msun)} & \colhead{(${\rm M_\odot L_\odot^{-1}}$)} & 
  \colhead{(keV cm$^2$)}  & \colhead{(keV cm$^2$)} & \colhead{} & \colhead{(kpc)} & \colhead{}
}
\startdata
NGC\thin 1332 & 11.9/8 & $0.52^{+0.41}_{-0.28}$ & $1.16^{+0.12}_{-0.14}$ & $0.55^{+0.31}_{-0.31}$ & $33.3^{+4.9}_{-4.6}$ & $0.969^{+0.12}_{-0.098}$ & \ldots & \ldots\\
          &    & (0.49) & (1.15) & (0.39) & (34.4) & (0.951) & \ldots & \ldots\\
NGC\thin 4261 & 22.9/12 & $0.44^{+0.28}_{-0.24}$ & $1.621^{+0.071}_{-0.092}$ & $0.586^{+0.12}_{-0.097}$ & $77.6^{+8.7}_{-8.2}$ & $1.325^{+0.066}_{-0.066}$ & $13.4^{+9.4}_{-3.5}$ & $0.47^{+0.21}_{-0.31}$\\
          &    & (0.48) & (1.627) & (0.558) & (78.6) & (1.332) & (13) & (0.54)\\
NGC\thin 4472 & 22.1/14 & $0.64^{+0.61}_{-0.33}$ & $1.473^{+0.085}_{-0.11}$ & $1.3^{+0.25}_{-0.37}$ & $66.1^{+11}_{-8.4}$ & $1.31^{+0.1}_{-0.13}$ & $5.23^{+1}_{-0.76}$ & $0.67^{+0.11}_{-0.16}$\\
          &    & (0.69) & (1.486) & (1.23) & (66.7) & (1.3) & (5.32) & (0.63)\\
\enddata
\tablecomments{The fit results for each galaxy, {adopting the \mbh-\sigmac\
prior for \mbh}. We show $\chi^2$/dof for 
the best-fit model (\S~\ref{sect_results}) and, for each interesting parameter, the most 
probable value and 90\%\ confidence regions, marginalized over the other parameters. 
Since the set of marginalized values
is not necessarily the set of parameter values which 
maximizes the posterior, we also list in parentheses the parameter values 
which correspond to such a ``best-fit''.
}
\end{deluxetable*}
The model described in \S~\ref{sect_models} was able to fit
the density and temperature profiles well and the best-fitting models are 
shown in Fig~\ref{fig_profiles}.  Strikingly, we do not 
see a sharp central temperature spike arising from the gravitational
influence of the black hole, similar to that previously reported for NGC\thin 4649.
There is a modest temperature rise in the centre of NGC\thin 1332, but this is
mostly a consequence of the peaked stellar density profile and is a feature
of our models even if \mbh$=0$. Of all the galaxies, only our model fits to 
NGC\thin 4472 show any evidence of a higher central temperature in the presence of
the black hole, and even then the effect is very weak, as shown in 
Fig~\ref{fig_n4472_profile}
(which compares the best-fitting temperature and density models with 
and without a central black hole). We discuss the reasons for this difference from
NGC\thin 4649 in \S~\ref{sect_mbh}.

The interesting marginalized parameters from our fit are given in 
Table~\ref{table_results}. For each object, the dark matter
halo mass and concentration
were degenerate with each other and we could not 
place interesting constraints on the virial mass (at least for
NGC\thin 1332 and NGC\thin 4472) without applying an 
additional prior, such as a restriction on the allowed baryon fraction
\citepalias[see][]{humphrey06a}. Rather than imposing such an ad hoc constraint, 
since neither of the dark matter halo parameters strongly correlate with 
\mbh\ or M/\lj, we simply marginalized over them and do not report them here.
For NGC\thin 1332, we were able to fit the data using 
a steep ($\beta_1 = 0.96$) entropy profile without any evidence of a break;
we could not constrain $s_{brk}$ and so we fixed it to a large value, 
outside the field of view. For NGC\thin 4261 and NGC\thin 4472 the entropy
profiles were similarly steep in the centre of the galaxy, but flattened 
outside $\sim$5--10~kpc. We show the entropy data-points, derived 
from the temperature and density data-points, plus the best-fitting entropy models
in Fig~\ref{fig_entropy_profiles}. This characteristic shape for the 
entropy profile is similar to that found in recent observations of X-ray bright galaxy groups
\citep{jetha07a,finoguenov07a,gastaldello07b,sun08a}, but $s_{brk}$ is at a 
comparatively smaller scale.

\begin{figure*}
\plotone{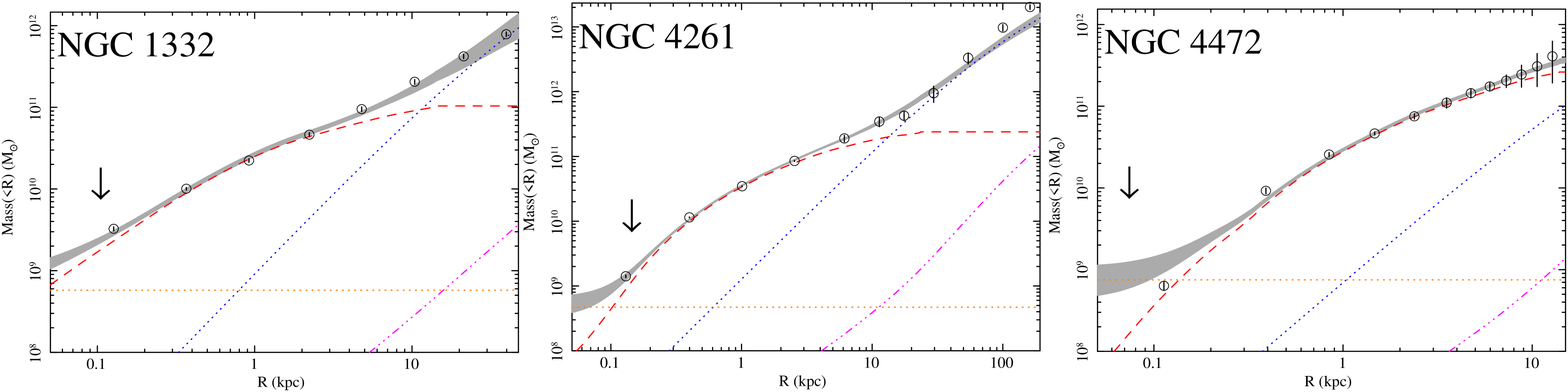}
\caption{Radial mass profiles for each galaxy, derived from our fits
to the \chandra\ data. The solid (black) lines indicate the total enclosed
mass, the dashed (red) lines indicate the stellar mass, the dotted
(blue) lines were the dark matter contribution, the dash-dot (orange) lines
are the black hole mass and the dash-dot-dot-dot (magenta) line 
is the gas mass contribution. The grey shaded regions indicate the 
1-$\sigma$ error on the total mass distribution. Overlaid are a 
set of data-points (and 1-$\sigma$ errors)
derived from a more ``traditional'' mass
analysis (\S~\ref{sect_nonparametric}), which generally agree very well
with the fitted models (we stress the models are {\em not} fitted to 
these data-points, but are derived separately). The arrows indicate the 
spatial scale corresponding to 1\arcsec.\label{fig_mass_profiles}}
\end{figure*}
The measured radial distribution of total gravitating mass is shown 
in Fig~\ref{fig_mass_profiles} for each galaxy,
along with the contribution from each of the separate mass components.
We found that the stellar mass component dominates 
within $\sim$5--10~kpc, with a significant contribution of dark matter 
required to explain our observations at larger radii in NGC\thin 1332
and NGC\thin 4261 (for NGC\thin 4472, the data we fitted did not extend
to large enough radii for the influence of the dark matter to be easily
discerned). These results are consistent with our previous analysis for
NGC\thin 4261 (and NGC\thin 4472), despite the much simpler stellar
mass profiles used in that work  \citepalias{humphrey06a}. 
\begin{figure*}
\plotone{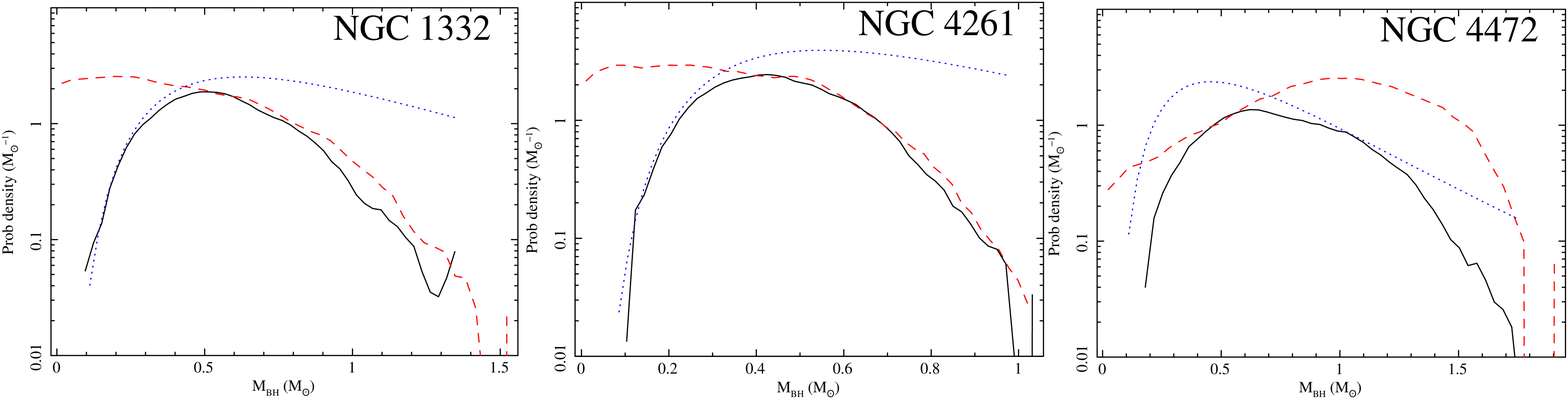}
\caption{{Marginalized probability density distributions for \mbh\ for each galaxy.
We show, as a solid line,
the probability density when the  \mbh-\sigmac\ relation was used as a 
prior. The dashed (red) lines are the probability density function with a flat prior
and the dotted (blue) lines are the \mbh-\sigmac\ relation 
prior itself; both have been 
arbitrarily scaled for clarity. Also for clarity, we smoothed the distribution functions with a fourth-order Savitzky-Golay
filter, which spans $\sim$20\%\ of the \mbh\ range for each system.
For NGC\thin 1332 and NGC\thin 4261, the shape of the
probability density function below $\sim3\times 10^8$\msun\ is mostly determined
by the prior, while above $\sim 4\times 10^8$\msun, the constraints are mostly
given by the X-ray data. The X-ray data alone provide only upper limits for
these two objects. For NGC\thin 4472, the shape of the probability density function is much less
sensitive to the prior.}
\label{fig_prob_density}}
\end{figure*}

\begin{figure*}
\plotone{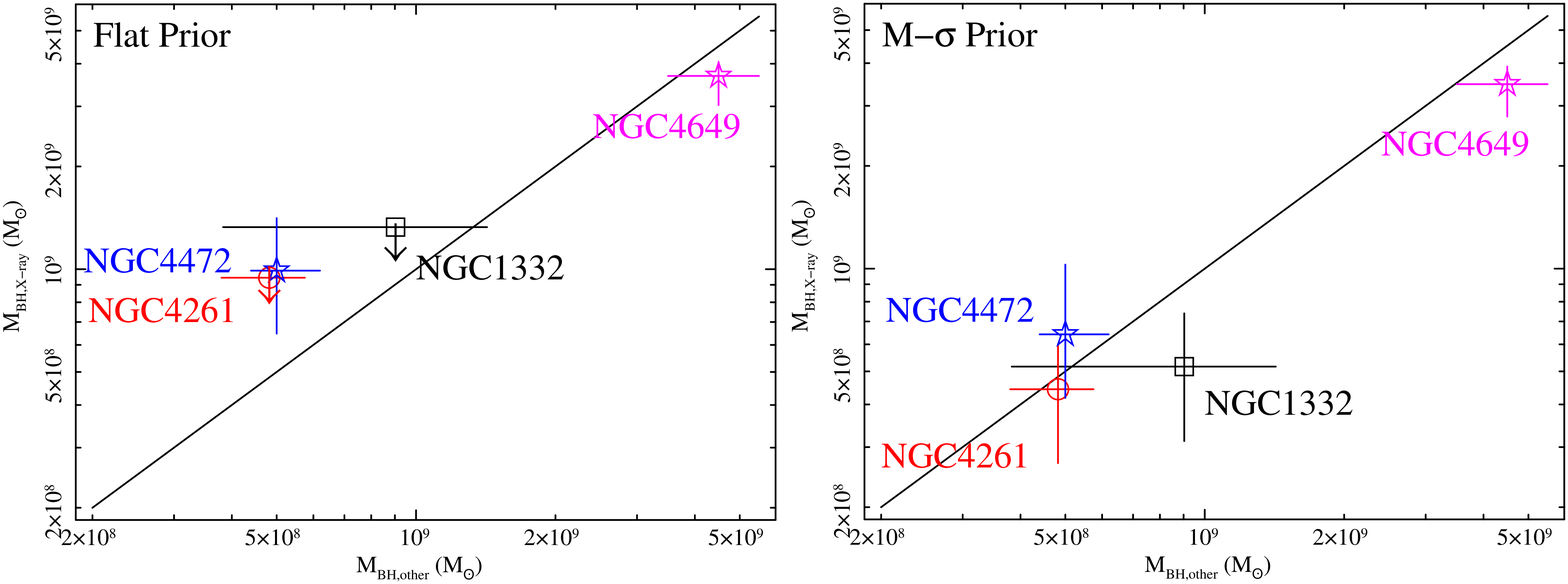}
\caption{{\em Left panel:} the derived black hole mass (\mbh) from the X-ray 
data, assuming a flat prior for \mbh, shown {\em versus}
black hole determinations from stellar dynamics (for NGC\thin 4472 and 
NGC\thin 4649; stars), gas disc dynamics (NGC\thin 4261; circle) or the M-$\sigma$
relation (NGC\thin 1332; square). {Error-bars correspond to 1-$\sigma$ uncertainties, although 
for NGC\thin 1332 and NGC\thin 4261 we show the 3-$\sigma$ upper limit
(marked as an arrow)}. The dotted line indicates $y=x$.
{\em Right panel:} The same, but using the 
\mbh-\sigmac\ relation as a prior on \mbh. Note that the 
comparison of the NGC\thin 1332 \mbh\ measurement using this prior with the 
\mbh-\sigmac\ relation, as shown in the right panel, 
should be treated with caution.
We find good agreement between
the masses measured with the different techniques.\label{fig_mbh}}
\end{figure*}
Since the stellar mass is non-negligible in the innermost bin, the constraints
on \mbh\ for these galaxies are less tight than we were able to obtain
for NGC\thin 4649. Unfortunately, this makes the results depend, to some degree,
on the choice of prior for \mbh. In Fig~\ref{fig_prob_density}, we compare the marginalized
probability density functions for \mbh\ for the case where the \mbh-\sigmac\ is used as a prior and 
for the flat prior case with the probability density function of the 
\mbh-\sigmac\ prior itself.
For NGC\thin 4472, there is a clear peak in the probability density when the 
flat prior is applied, allowing \mbh\ to be constrained to 
\mbh=$(1.04^{+0.62}_{-0.76})\times 10^9$\msun\ (90\%\ confidence), and excluding \mbh=0 at
98\%\ probability (or ``2.3$\sigma$'')\footnote{This result is also obtained with a 
simple frequentist analysis: the difference in $\chi^2$ between the best-fitting
model and the best-fitting model with \mbh=0 is 5.5 (which corresponds to a 98\%\
confidence region). On the basis of an f-test, the significance of the black hole 
detection is therefore $\sim$94\%.}. Applying the \mbh-\sigmac\ relation as a prior causes the 
probability density peak to shift and narrow somewhat, but our conclusions are not
qualitatively changed.
For NGC\thin 1332 and NGC\thin 4261 the 
prior has a more significant effect on the results, however;
below $\sim 3\times 10^8$\msun, the (\mbh-\sigmac) prior completely dominates the shape of the 
probability density function, while the X-ray data largely determine its shape above 
$\sim$4$\times 10^8$\msun. Put another way, the X-ray data alone provide a firm
upper limit for \mbh\ for these two galaxies, while
the \mbh-\sigmac\ relation eliminates the portion of parameter space
in which \mbh$\rightarrow$0, which is inconsistent with the apparent ubiquity of 
SMBHs in giant early-type galaxies.

In Fig~\ref{fig_mbh} we compare our constraints on \mbh\ for each galaxy
with measurements using other methods. 
For NGC\thin 4472, we compare to the 
stellar dynamical measurement of  \citet[][omitting a dark matter halo]{gebhardt09a}, for NGC\thin 4261 we 
use the gas disc dynamics measurement of \citet{ferrarese96a} and 
for NGC\thin 1332, for which there is no published dynamical \mbh\ measurement,
we simply compare to \mbh\ estimated from the M-$\sigma$ relation. We also
include the results for a re-analysis of NGC\thin 4649, applying the same Bayesian
methods and stellar light modelling used in the present paper
\citep[and the stellar dynamical analysis of][]{shen09a}.
For completeness, we show the results using both choices of prior on \mbh. 
As expected, for NGC\thin 4472 and NGC\thin 4649, the constraints are relatively
insensitive to this choice. 
There is clearly agreement between the X-ray and stellar dynamical mass measurements.
Particularly striking is that, for NGC\thin 4649, 
the error-bars obtained from the X-ray method are roughly half the size of those
found by the stellar dynamical modelling. The X-ray measurement
for NGC\thin 4472 is marginally higher than the dynamically-determined \mbh. However,
in the revised analysis of NGC\thin 4649 by
\citet{shen09a}, \mbh\ increased by a factor $\sim$2 from that obtained by 
\citet{gebhardt03a}.  Whether this increase is due to inclusion of a dark matter halo in
the newer models models or improved
dynamical modeling will require additional studies. Nevertheless, it is plausible
that there could be a similar revision for NGC~4472 when the dark matter halo is included, 
which would bring it into even better agreement with the X-ray data.

For NGC\thin 4261, the data unambiguously rule out that 
\mbh\ determined from X-rays
substantially exceeds the gas disk dynamical measurement. Although the lower limit from 
our analysis is largely determined by the prior (so we cannot place stringent limits
on any bias due to nonthermal pressure support, although see \S~\ref{sect_mbh}), there
is no evidence from our analysis of a discrepancy between the gas disk dynamics
measurement and our X-ray results.
For NGC\thin 1332, our results constitute the first direct measurement of the SMBH
mass, and so in Fig~\ref{fig_mbh} we compare it to the \mbh-\sigmac\ relation
rather than a measurement derived from stellar dynamics. We find consistency; the 
X-ray determined \mbh\ clearly does not lie significantly above this relation.
\begin{deluxetable*}{lrrrrrr}
\centering
\tablecaption{Systematic error budget\label{table_syserr}}
\tablehead{
\colhead{Test} & \multicolumn{2}{c}{NGC\thin 1332} & \multicolumn{2}{c}{NGC\thin 4261} & \multicolumn{2}{c}{NGC\thin 4472} \\
 & \colhead{\mbh} & \colhead{$M/L_J$} & \colhead{\mbh} & \colhead{$M/L_J$} & \colhead{\mbh} & \colhead{$M/L_J$}\\
\colhead{} & \colhead{($10^9$\msun)} & \colhead{(${\rm M_\odot L_\odot^{-1}}$)} & \colhead{($10^9$\msun)} & \colhead{(${\rm M_\odot L_\odot^{-1}}$)} & \colhead{($10^9$\msun)} & \colhead{(${\rm M_\odot L_\odot^{-1}}$)}
}
\startdata
Most probable & $0.52^{+0.41}_{-0.28}$& $1.16^{+0.12}_{-0.14}$& $0.44^{+0.28}_{-0.24}$& $1.62^{+0.07}_{-0.09}$& $0.64^{+0.61}_{-0.33}$& $1.47^{+0.09}_{-0.11}$\\\hline
$\Delta$Prior(entropy) & $\pm 0.04$ & $\pm 0.02$& $^{+0.09}_{-0.04}$ & $-0.01$& $+0.17$ & $-0.02$\\
$\Delta$Prior(M/L)& $-0.04$ & $-0.03$& $-0.13$ & $+0.07$& $+0.05$ & $-0.02$\\
$\Delta$Prior(\mvir) & $+0.11$ & $-0.01$& $+0.04$ & $+0.00$& $+0.24$ & $-0.02$\\
$\Delta$Prior(Gultekin)& $-0.04$ & $-0.03$& $-0.10$ & $+0.00$& $+0.10$ & $+0.01$\\
$\Delta$Prior(Flat \mbh)& $-0.27$ & $+0.00$& $-0.16$ & $+0.01$& $+0.35$ & $-0.02$\\
$\Delta$Stars(geometry)& \ldots & \ldots& $-0.09$ & $-0.19$& $^{+0.08}_{-0.01}$ & $-0.11$\\
$\Delta$Stars(deprojection)& \ldots & \ldots& $-0.02$ & $-0.14$& $-0.22$ & $-0.24$\\
$\Delta$Stars(Sersic)& $-0.09$ & $-0.13$& $-0.22$ & $-0.06$& $-0.15$ & $-0.15$\\
$\Delta$DM & $-0.09$ & $+0.07$& $-0.07$ & $+0.10$& $+0.07$ & $+0.09$\\
$\Delta$Background & $+0.09$ & $-0.14$& $+0.02$ & $-0.06$& $-0.16$ & $+0.08$\\
$\Delta$\nh & $-0.07$ & $+0.03$& $-0.03$ & $+0.01$& $\pm 0.03$ & $^{+0.03}_{-0.02}$\\
$\Delta$XRB & $-0.10$ & $+0.04$& $\pm 0.02$ & $-0.02$& $-0.15$ & $\pm 0.02$\\
$\Delta$Plasma Code& $-0.08$ & $+0.03$& $-0.01$ & $-0.06$& $+0.25$ & $+0.04$\\
$\Delta$Bandwidth& $-0.11$ & $\pm 0.03$& $-0.04$ & $-0.02$& $^{+0.33}_{-0.16}$ & $\pm 0.07$\\
$\Delta$Statistic& $-0.06$ & $-0.01$& $-0.06$ & $+0.00$& $+0.01$ & $-0.05$\\
$\Delta$Centroid& $+0.08$ & $-0.12$& \ldots & \ldots& $+0.02$ & $+0.00$\\
$\Delta$Distance & $\pm 0.02$ & $\pm 0.09$& $-0.07$ & $^{+0.17}_{-0.15}$& $-0.03$ & $\pm 0.06$\\
$\Delta$Rotation& $-0.08$ & $+0.09$& $-0.05$ & $+0.01$& $+0.01$ & $-0.01$\\
\enddata
\tablecomments{Estimate of the likely impact of potential sources of systematic errors
in our analysis on our measurements of \mbh\ and the stellar 
M/\lj. We give the marginalized value and 90\%\ statistical errors for 
each galaxy (top line). For a series of tests (see text), the change in the 
marginalized value caused by an arbitrary change of the underlying assumptions in our
analysis is given.
We stress that these systematic errors 
should {\em not} be added in quadrature with the statistical errors. In most (but not
all) cases, the systematic errors are no larger than the statistical errors.}
\end{deluxetable*}
\section{Systematic error budget} \label{sect_systematics}
In this section we address the sensitivity of our results
to various data-analysis choices which were made, including
the choice of prior. In most cases it is difficult
or impractical to express these assumptions through a single
additional model parameter over which one might hope to 
marginalize, and so we adopt the pragmatic approach of 
investigating how our results change if the assumptions
are arbitrarily adjusted. We focused on those systematic
effects likely to have the greatest impact on our conclusions,
and list in Table~\ref{table_syserr} how the ``most probable'',
marginalized \mbh\ and M/\lj\ measurements for each galaxy are 
affected by each test. These estimates constitute a likely
upper limit on the magnitude of each systematic uncertainty and
we stress they {\em should not be added in quadrature with the statistical errors}.
We outline below how
each test was performed, so  those readers uninterested in
the technical details of our analysis may wish to proceed
directly to \S~\ref{sect_discussion}.
\subsection{Priors} \label{sect_systematics_priors}
As outlined in \S~\ref{sect_prior}, the choice of priors on several of the
parameters are  arbitrary. We therefore considered the 
effect of replacing each arbitrary choice with another, reasonable
prior. Specifically, for each parameter describing the entropy
profile, and for the central gas pressure, we switched
from a flat prior on the parameter to a flat prior on its logarithm.
The effect, summarized in Table~\ref{table_syserr} as $\Delta$Prior(entropy),
is typically smaller than the statistical errors. Similarly, we have
experimented with replacing the flat prior on M/\lj\ with a Gaussian
prior, the mean and sigma of which is determined from the stellar
population synthesis models; see \S~\ref{sect_mass_to_light}.
The effect of this choice is summarized as $\Delta$Prior(M/L) in 
Table~\ref{table_syserr}.
To assess the importance of the \mvir\ prior,
we experimented with instead fixing the \mvir\ to a canonical
value, based on the  empirical relation of 
\citet{conroy08a}, which links the virial mass of a halo to the stellar mass
it hosts. We computed the stellar mass from the \lj\ values 
given in Table~\ref{table_sample}
and the best-fitting stellar M/\lj\ ratio for each galaxy. The effect of 
this choice is summarized as $\Delta$Prior(\mvir).
{Finally, we investigated the effect of our \mbh\ prior by adopting
the revised and updated \mbh-\sigmac\ relation
for early-type galaxies reported by 
\citet{gultekin09a} as the prior. The results ($\Delta$Prior(Gultekin))
indicate that this has little effect on our conclusions. In addition,
we also report the result of using the (poorly motivated) 
flat prior on \mbh ($\Delta$Prior(Flat \mbh)),
as discussed in \S~\ref{sect_prior}.
In general, adopting the flat prior had a larger impact on the derived parameters 
than any of the other choices.}
\subsection{Stellar model} \label{sect_systematics_stars}
As shown in \citetalias{humphrey06a}, careful modelling of the stellar 
light is necessary to obtain an accurate measurement of the mass-to-light
ratio. Similarly, since we did not typically resolve scales at which the 
black hole is completely dominating the gravitating mass, we also expect 
\mbh\ to be sensitive to this modelling. For NGC\thin 4261 and NGC\thin 4472,
both of which have uncertain geometry, we experimented with 
using a prolate model instead of an oblate model, and changing the inclination
from 90$^\circ$ to 45$^\circ$. In NGC\thin 1332, the
geometry is less uncertain since the presence of the disc allowed us 
constrain the inclination and symmetry axis. {The effect of these 
choices is listed as $\Delta$Stars(geometry) in Table~\ref{table_syserr}.

In \S~\ref{sect_stars} we compared our deprojected stellar density profiles to 
those obtained by using the deprojection method of \citet{gebhardt96a}, finding 
general consistency within the central $\sim$10~kpc. Nevertheless, to quantify
any systematic uncertainties associated with the deprojection method, we also used 
the V-band profiles for NGC\thin 4261 and NGC\thin 4472. Given the different
filters, we corrected the M/L ratios obtained based on the colours measured in 
a $\sim$30\arcsec\ aperture, taking the V-band data from \citet{sandage78a}
and computing the J-band magnitudes from the \twomass\ data discussed in \S~\ref{sect_stars}.
We found that use of this profile did not lead to significantly different conclusions ($\Delta$Stars(deprojection) in Table~\ref{table_syserr}).
For all three galaxies, we also
investigated the effect of adopting a simple Sersic approximation for the 
stellar light, as shown in Fig~\ref{fig_optical_images}. This is useful as residual 
dust extinction in the centre of NGC\thin 4261 likely leads us to underestimate the 
stellar density, so that the true density profile lies intermediate between the 
preferred deprojected profile and the simple Sersic parameterization.
The impact of this test is listed as $\Delta$Stars(sersic) in 
Table~\ref{table_syserr}.
All three tests had a noticeable effect on the best-fitting results,
comparable to the measured statistical errors. Nevertheless, these sources
of uncertainty should not  qualitatively affect our conclusions.}
\subsection{Dark matter halo} \label{sect_syserr_dm}
{Although we showed in \citetalias{humphrey06a} that the dark matter halos of 
a sample of early-type galaxies were consistent with NFW profiles obeying the 
concentration-virial mass relation predicted by numerical structure formation 
simulations \citep[see also][]{buote07a}, in some recent stellar dynamical analyses
it has been argued that a ``cored logarithmic'' profile may provide a better fit
\citep[\eg][]{thomas07a}.
Therefore, to test the extent to which the modelling of the dark halo affects our results,
we experimented with
using the cored logarithmic profile in place of NFW. We found this had a slight
effect (summarized as $\Delta$DM in Table~\ref{table_syserr}), 
comparable to the other sources of error. We will return to the question of 
the optimal shape of the dark matter halo in a future paper.}
\subsection{Spectral modelling}
We investigated several potential sources of systematic uncertainty in our spectral
analysis. First, we considered the impact of our background treatment, by adopting
the spectra derived from the standard background template events files distributed with the
\ciao\ \caldb, having renormalized them to match the count-rate in the 9--12~keV
band. This choice had a measurable effect on our results, comparable to the statistical
uncertainties ($\Delta$Background in Table~\ref{table_syserr}). We also investigated 
how our results are affected by changing \nh\ by $\pm$25\%\ ($\Delta$\nh), replacing the
bremsstrahlung component with a $\Gamma$=1.5 powerlaw or adjusting its temperature by 
$\pm$25\%\ ($\Delta$XRB), using a MEKAL rather than an APEC plasma ($\Delta$Plasma Code),
or changing the energy range over which the fitting was performed to
0.4--7.0~keV, 0.5--2.0~keV or 0.7--7.0~keV ($\Delta$Bandwidth). Two further tests we
carried out were using the $\chi^2$ fit statistic rather than the C-statistic
($\Delta$Statistic), which had only
a very small effect, and moving the centroid of the extraction regions by up to
$\sim$0.5\arcsec, consistent with the accuracy we estimate for our centroiding procedure
($\Delta$Centroid).
This latter test was not carried out for NGC\thin 4261, since the X-ray detection
of the AGN makes the centroiding very accurate.
\subsection{Other tests}
Two final issues we investigated were the uncertainty in the  adopted 
distance to the galaxies, and the 
possibility of gas rotation. For the former, we adjusted the distance by the 
1-$\sigma$ errors reported in \citet{tonry01} and the effect is summarized as 
$\Delta$Distance in Table~\ref{table_syserr}. To test the impact of the latter, we 
allowed the gas to rotate along with the stars and included an additional nonthermal 
pressure component to the equation of hydrostatic equilibrium, following \citet{fang09a}.
We consider this to be likely an upper limit on the gas rotation; in none of the galaxies
do we find the X-ray isophotes are much flatter than the optical light, as might be
expected if the gas is rapidly rotating
\citep[for a more detailed discussion, see][]{brighenti09a}. 
For NGC\thin 1332 we adopted the major axis rotation profile from \citet{dressler83a},
for NGC\thin 4261 we used the minor axis profile from \citet{bender94a} and for NGC\thin 4472
we used the major axis profile of  \citet{sanchezblazquez07a}. Assuming that the gas
rotates as a rigid body, we find (unsurprisingly) that rotation 
increases the amount of enclosed mass we measure, but the effect ($\Delta$Rotation) 
is  not larger than the statistical errors on M/\lj\ or \mbh.

\section{Discussion} \label{sect_discussion}
\subsection{Black hole mass measurements} \label{sect_mbh}
Under the assumptions of hydrostatic equilibrium, spherical symmetry, 
a single-phase ISM and a stellar mass component which follows the optical light,
we have constrained the masses of the SMBHs at the 
centres of three galaxies. Combined with our previous measurement for NGC\thin 4649
\citepalias{humphrey08a}, this demonstrates that hydrostatic X-ray methods are a 
practical, competitive means for SMBH mass determination.
We found that the masses determined from the X-ray method
are in good agreement with those found from other techniques--- specifically
stellar kinematics for NGC\thin 4472 and NGC\thin 4649 \citep{gebhardt09a,gebhardt03a},
gas disc dynamics for NGC\thin 4261 \citep{ferrarese96a}, and the
\mbh-\sigmac\ relation for NGC\thin 1332. This makes NGC\thin 4261, NGC\thin 4472
and NGC\thin 4649 three of only a small handful of galaxies which have \mbh\
determined by more than one method.

{The agreement between our results and the SMBH masses obtained from other methods (especially for NGC\thin 4472 and NGC\thin 4649, for which the constraints
do not strongly depend on the prior), provides support for the assumptions in our analysis.
Key among these assumptions is that of hydrostatic equilibrium. 
While one might expect deviations from this approximation to be most 
prevalent close to the SMBH, since the ejecta from accretion episodes are
known to interact with the the ISM at large scales, the local freefall
timescale is shortest in the central region of the galaxy so that 
hydrostatic equilibrium
will quickly be re-established if the gas is stirred up. 
Furthermore, the physics of the 
coupling between an AGN and the surrounding ISM remains very uncertain, particularly on such small scales.
We would expect deviations from hydrostatic equilibrium most likely
to manifest themselves in nonthermal pressure support
\citep[\eg][]{churazov08a,zappacosta06a}, leading the X-ray method to underestimate the 
true \mbh. As is clear from Fig~\ref{fig_mbh}, the 
X-ray measurement for NGC\thin 4472 is slightly
{\em higher} than the measurement from stellar dynamics, while those for NGC\thin 4649 are very
close, suggesting that any such nonthermal pressure is small.
While, for NGC\thin 1332 and NGC\thin 4261, there is also no evidence that the 
X-ray determined mass is systematically underestimated
(since the upper limits imposed 
by the X-ray data are consistent with the gas disk measurement for NGC\thin 4261
and the \mbh-\sigmac\ relation for NGC\thin 1332), the data do not allow us to 
rule this out, since the lower limit on \mbh\ is 
largely determined by the prior. To investigate this in more detail
there is a clear need for higher-quality \chandra\ data,
which should allow more stringent upper limits to be placed on  \mbh.
Since the \mbh-\sigmac\ relation is not sufficiently accurate for this purpose,
this will also require an accurate \mbh\ determination from stellar dynamics
for NGC\thin 1332. Despite these concerns for these two systems, 
it is unlikely that any deviations from
hydrostatic equilibrium are restricted solely to the central bin, and deviations at larger
radius should produce residuals in our fit to the temperature and density profiles
and a systematically under-estimated stellar M/\lj\ ratio. As we discuss in \S~\ref{sect_he},
the good fits to these profiles at larger radii, combined with reasonable M/\lj\ ratios for
each galaxy in fact suggest that hydrostatic equilibrium is a reasonable approximation.

Another crucial assumption in our analysis is that the ISM is approximately
single phase at any radius. Although this approximation is not exact for NGC\thin 4472,
as evinced by regions of cool gas in its centre \citep{caon00a}, the good overall agreement
between \mbh\ determined from the X-rays and optical methods strongly suggests
 that it is, nevertheless,
fairly accurate (implying a small filling factor for the cool phase). 
The dusty central discs in NGC\thin 1332 and NGC\thin 4261 similarly 
reveal departures from a perfectly single phase ISM. While, as discussed above,
 we cannot rule out the possibility
that the X-rays may underestimate the mass for these two systems, provided the filling factor
of the cool phase is small (for example, if the gas is restricted to this disc-like geometry),
the density and pressure of the hot phase obtained from our X-ray measurements will be 
fairly accurate, and thus sufficiently accurate for our
purposes.

The agreement we found between the \mbh\ measurements 
not only provides strong support for our X-ray approach
but also crucial, independent verification that the dynamical \mbh\ measurement techniques are
accurate. As discussed in \S~\ref{sect_results}, the stellar dynamical measurements 
for NGC\thin 4472 and NGC\thin 4649 are marginally lower than \mbh\ obtained from the 
X-rays, but this could arise  from the omission of the dark matter halo
in the stellar modelling \citep[\eg][]{shen09a}. Notwithstanding, this comparison 
suggests that
deviations from triaxiality, uncertainties in the inclination of the galaxy and 
pathological orbital structure cannot give rise to additional systematic 
errors larger than a factor $\sim$2 in \mbh\ for either galaxy (unless they are compensated for
by finely tuned systematic errors in the X-ray determined mass).
The similar agreement for NGC\thin 4261 also suggests that any systematic errors in the 
\mbh\ determination are not significantly larger than the current statistical errors.
This is interesting since \citet{ferrarese96a} used sparsely sampled HST FOS data,
while higher-quality STIS data suggest that non-Keplerian motions may be non-negligible
\citep{noelstorr03a}.
Since the prior dominates the probability density function at low masses the agreement we
find partially reflects the fair agreement between the gas disk measurement of NGC\thin 4261 
and the \mbh-\sigmac\ relation of \citet[][]{tremaine02a}, which was derived in part
from this data-point. Still, it is very unlikely that this one point strongly 
influenced either the shape or scatter of their fit, so, 
combined with the tight upper
limit on the mass from the X-ray data--- which is also consistent with this datum--- the
agreement is very encouraging.}

In NGC\thin 4649, the primary effect of the SMBH on the gas was to introduce
a sharp central temperature peak, similar to that predicted by \citet{brighenti99c}. 
However, none of the three galaxies in the present study shows such a 
pronounced effect. This difference in behaviour can be understood in part 
as a consequence of the black hole mass. 
In NGC\thin 4649, the SMBH dominates the mass profile
for much of the region enclosed by the central bin. Technically, 
the stellar mass enclosed within this
bin is larger, but what is more important is the mass evaluated not at the 
outer radius of the bin, but at an appropriate intermediate position
\citep[see \eg][]{gastaldello07a}. Using an appropriate radius,
the black hole actually dominates the mass, contributing $77^{+3}_{-8}$\%\ of the 
total   (see Fig~\ref{fig_mass_profiles}), whereas in the other
galaxies \mbh\ is almost an order of magnitude lower and the stellar mass
largely dominates (formally, it contributes $18\pm10$\%, $35\pm15$\%, and
$65^{+10}_{-22}$\%\ of the mass in the central bin 
for NGC\thin 1332, NGC\thin 4261 and NGC\thin 4472, respectively).
Correspondingly, therefore, the SMBH has a smaller
influence on the gas. 
Another key difference between NGC\thin 4649 and the other galaxies is the
shape of its entropy profile, which is very flat within $\sim$500pc,
in contrast to the approximately $s\propto R$ profiles at small scales 
shown in Fig~\ref{fig_entropy_profiles}. Since a lower central entropy
tends to reduce the temperature of the gas in the centre, any temperature
peak is suppressed in NGC\thin 4472, NGC\thin 4261 and NGC\thin 1332.

One possible concern with our analysis is the dependence of the best-fitting
\mbh\ measurements on the \mbh-\sigmac\ relation, which we used as a prior.
In particular the \mbh-\sigmac\ relation is determined in large part from 
stellar dynamical measurements which do
not include dark matter halos. {Since this omission is able to bias the
measured \mbh, the exact shape or scatter in this relation could be in error, 
particularly at the critical high-mass end, which may have an impact on our measured
parameters. However, if significant, we would expect this to revise upwards the mass
expected from the prior, which would serve only to restrict further the parameter
space allowed (Fig~\ref{fig_prob_density}) and shrink our error-bars. 
Moreover, we obtained qualitatively the same results when the flat prior was used
in place of the \mbh-\sigmac\ relation (albeit we then only obtained upper limits
on \mbh\ for two of the four galaxies), giving us confidence that subtle revisions
to the \mbh-\sigmac\ relation will not change our conclusions.
Still, for a precision comparison of \mbh\ determined by different techniques,
and, in particular, constraining the \mbh-\sigmac\ relation itself from X-ray 
data alone, deeper \chandra\ observations will be useful as higher-quality 
data generally result in reduced sensitivity to the priors.}

Although we have demonstrated that hydrostatic X-ray mass analysis is 
 potentially a
powerful means to determine \mbh\ in gas-rich early-type galaxies, such measurements
are contingent upon spatially resolved X-ray spectroscopy at the 
smallest possible scales. Given the limitations
of \chandra, it seems likely that this technique will only be applicable to a handful
more galaxies in the near future (\S~\ref{sect_targets}). 
However, with the spatial resolution promised by 
future X-ray missions, in particular \genx\ \citep{windhorst06a}, such measurements should
become routine. A factor $\sim$5 improvement in PSF would 
likely correspond to $\sim$2 orders of magnitude increase in
surveyable volume and hence number of accessible systems. 
Further, with such spatial resolution, it will become possible routinely to resolve
the sphere of influence of the SMBH, making the \mbh\ measurement no
longer dependent primarily on one data bin.

\subsection{Stellar M/\lj\ ratios} \label{sect_mass_to_light}
\begin{figure}
\includegraphics[width=3.2in]{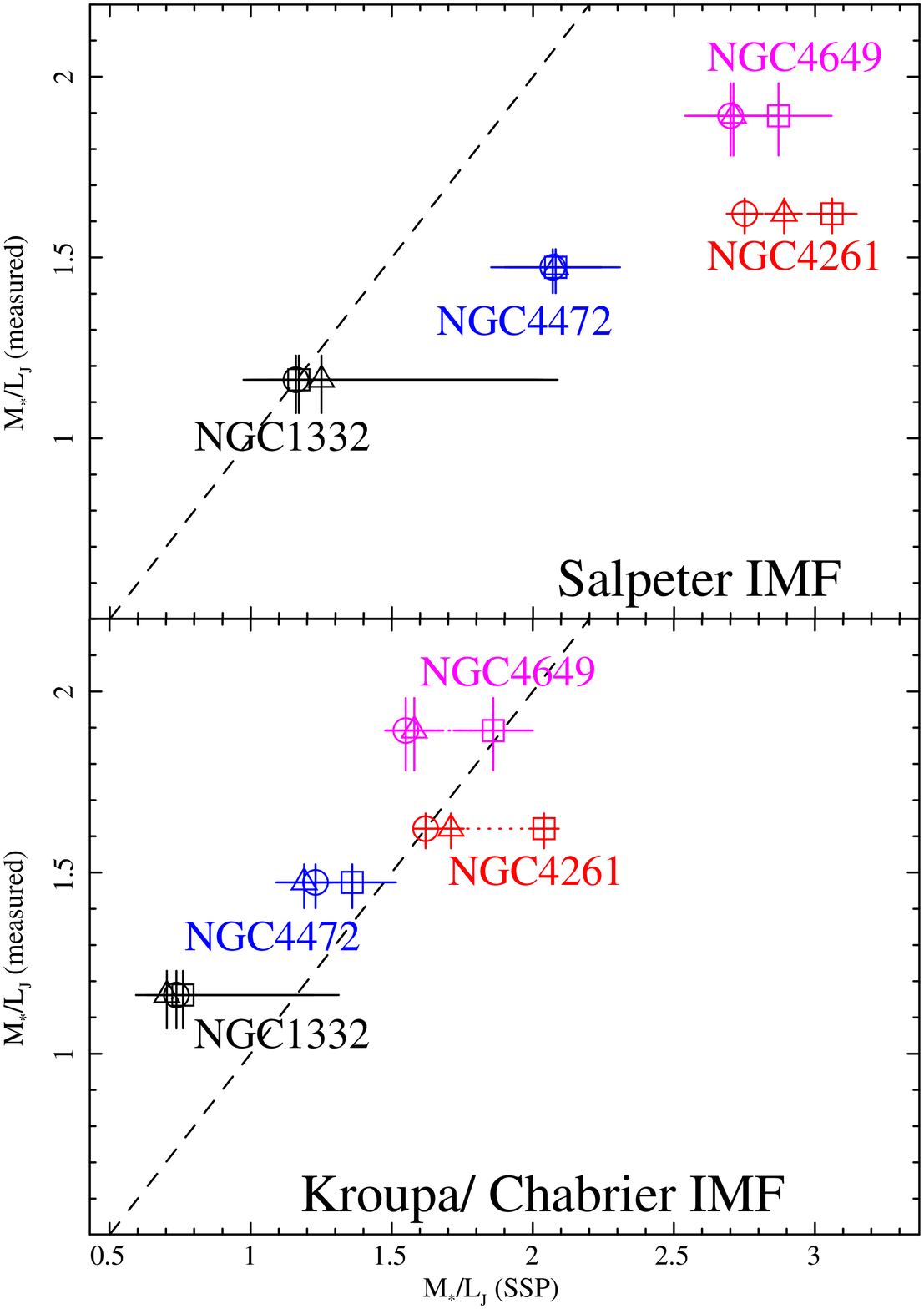}
\caption{{Comparison of the {\em stellar} M/\lj\ ratios determined from our 
X-ray analysis to the predictions of a selection of single burst 
stellar population (SSP) models. Squares denote the SSP models of 
\citet{maraston05a}, circles denote PEGASE 
\citep{fioc97a} models and triangles denote the models of \citet{bruzual03a}. 
In the upper panel we show the synthetic SSP M/\lj\ 
ratios computed for a Salpeter IMF and in the lower panel, we use the \citet{kroupa01a}
IMF for the \citeauthor{maraston05a} and PEGASE models, and the \citet{chabrier03a}
IMF for the \citeauthor{bruzual03a} models. The dotted line denotes the $y=x$ relation.
We find good overall agreement between the measured M/\lj\ ratios
 and the models using a Kroupa or Chabrier IMF, suggesting that the 
gas is close to hydrostatic.}\label{fig_mass_to_light}}
\end{figure}
{In addition to the \mbh\ measurements, our hydrostatic models also enabled
us to constrain the stellar M/\lj\ ratios, under the assumption that the 
stellar mass follows light. In Fig~\ref{fig_mass_to_light}, we compare
our measured values to those predicted by three sets of single-burst stellar population (SSP)
models, given the stellar age and abundance
listed in Table~\ref{table_sample}. In particular, we used the models 
of \citet[][using the  updated model grids made available by the 
author\footnote{{http://www.dsg.port.ac.uk/$\sim$marastonc}. At the metallicity of these galaxies, the differences between the models with a blue and 
red horizontal branch are negligible.}]{maraston05a}, the PEGASE code version 2 \citep{fioc97a} and the 
models of \citet{bruzual03a}. With the \citeauthor{maraston05a} and PEGASE models, we
investigated both the Salpeter initial stellar mass function (IMF)
and that of \citet{kroupa01a}, while for 
the \citeauthor{bruzual03a} models we used the Salpeter and the \citet{chabrier03a} IMFs,
and the ``Padova 1994'' evolutionary tracks.
To compute the predictions of the SSP models, we linearly interpolated 
\mstars/\lj\ from the synthetic values tabulated on a convenient grid as a function
of metallicity and age. The observed M/\lj\ we reported in 
 \citetalias{humphrey08a} for NGC\thin 4649 is $\sim$20\%\ 
lower than shown here, which arises
due to the improved modelling of the stellar light in the present paper;
the galaxy was assumed to be spherical in \citetalias{humphrey08a}, whereas
it is appreciably flattened outside a few kpc \citep{peletier90a}, which
we take into account with the deprojection method outlined in \S~\ref{sect_stars}.

Although there is considerable scatter in the synthetic M/\lj\ values 
predicted by the SSP models, we find good overall agreement between the 
predictions of the models using a \citeauthor{kroupa01a} or \citeauthor{chabrier03a}
IMF and the results of our fits. In contrast, the models using the Salpeter IMF 
dramatically over-estimate M/\lj. 
While this general trend is consistent with
the conclusions of our previous work, in \citetalias{humphrey06a} our measured 
M/L ratios were typically significantly smaller even than those assuming a 
\citeauthor{kroupa01a} IMF.
In particular, for NGC\thin 4472, we  measured M/\lk\ to 
underestimate the SSP value by $\sim$40\%. The origin of this 
discrepancy, which has vanished in our updated analysis, can be easily traced
to the simpler form for the stellar density profile used in our earlier work.
In \citetalias{humphrey06a}, we modelled the stellar light as a
\citet{hernquist90} profile, 
which significantly differs from the density profile
derived in \S~\ref{sect_stars} outside $\sim$4~kpc. This difference was 
exacerbated by our not choosing the parameters for the stellar model in 
a self-consistent manner, thus introducing considerable additional systematic
error into our older M/\lk\ measurement (see \S~7.2 of \citetalias{humphrey06a}).

While the agreement in the lower panel of Fig~\ref{fig_mass_to_light} is striking,
it is unlikely that real early-type galaxies are monolithic, single-burst stellar populations.
Indeed, the presence of line-strength and colour gradients 
\citep[\eg][]{peletier90a,trager00a} indicates that the metallicity may vary radially and
at least some elliptical galaxies host multiple-aged stellar populations 
\citep[\eg][]{rembold05a}. Fitting SSP models to the Lick indices of such a system
in order to determine a mean age and metallicity may yield a biased result, although if 
one population dominates the stellar light, the bias may be small. Such a bias is most
problematical for the age, which is the most important factor influencing M/\lj\ for an
old system. Older populations typically have larger M/\lj\ ratios and so,
since the mean stellar ages we estimated for NGC\thin 4649 and NGC\thin 4261 are both
\gtsim 13~Gyr, the ``true'' M/\lj\ for these systems cannot be higher than our 
SSP predictions, for a given IMF. While this could in principle 
allow the X-ray data to be reconciled
with a Salpeter IMF, for these two systems this would require the stellar light to be 
largely dominated by stars aged less than $\sim$8~Gyr, implying a very large bias on the 
measured age. Given the good agreement for all the systems 
between the measured M/\lj\ ratios and the
SSP models computed for a  \citeauthor{kroupa01a} or \citeauthor{chabrier03a} IMF, a simpler
hypothesis is that the bias is actually small and the Salpeter IMF is not correct.

If the stellar population properties do vary radially, this can cause a violation of the 
assumption that the stellar mass strictly follows the light. Nevertheless,
in the very red photometric band
we adopt here the effect of varying the metallicity is very slight
(\eg\ significantly changing the metallicity of a $\sim$10~Gyr population, assuming a \citealt{kroupa01a}
IMF and the SSP models of \citealt{maraston05a}, from [Fe/H]=0.35 to -0.33 causes only
a $\sim$6\%\ change in the stellar M/\lj\ ratio), so a metallicity gradient is not a concern.
Furthermore, if an age gradient is present which might 
significantly affect the M/L ratio, it is likely to be give rise to 
a substantial colour gradient.  In practice, all four galaxies show only 
very slight evidence for any such gradient, at least in V-I colour
\citep[][see also the good agreement between the shapes of the 
deprojected V-band and J-band profiles shown in Fig~\ref{fig_optical_images}
for NGC\thin 4261 and NGC\thin 4472]{poulain88a,poulain94a}.
Therefore, the assumption that stellar mass follows J-band 
light is likely to be reasonable.

We note that
simple models of galaxy formation suggest that the condensation of gas into stars
should produce a gravitational reaction on the dark matter halo, leading to a 
cuspier dark matter profile through a process termed ``adiabatic contraction''
\citep{blumenthal86a,gnedin04a}. Such models increase the total gravitating 
mass in the centre of the galaxy substantially, thus lowering the 
stellar mass-to-light ratio one measures for a given object \citepalias{humphrey06a} 
when the models are applied. Our data clearly leave little room for a substantial 
reduction in the measured M/\lj\ ratios if we require consistency with SSP models.
This is in clear contradiction with the adiabatic contraction models as currently implemented, 
unless they operate in a finely tuned conspiracy with nonthermal pressure
(\S~\ref{sect_he}),
an IMF radically different from those discussed here, or a mixture of multiple-aged
stellar populations in each galaxy. This result is not surprising since similar difficulties
for the adiabatic contraction models have been reported for disk galaxies
\citep{dutton07a,gnedin07a} and recent simulations have found baryonic condensation
to a have a significantly smaller effect on the dark matter halo \citep{abadi09a}.
}

Finally, it is interesting to compare our measured stellar mass to light ratios with those obtained
from stellar dynamical methods. Unfortunately, most dynamical studies to date have involved
significant simplifications, such as omitting a dark matter component, assuming spherical
symmetry or limiting the generality of the orbital structure, which may prevent a fair 
comparison with our work. Of the those incorporating a dark matter halo, 
\citet{kronawitter00a} constructed spherical dynamical models for NGC\thin\ 4472,
but restricted the phase space distribution by applying an expansion using a set of  known
basis functions. They found M/L$_{B}\sim$8--10, correcting to our
adopted distance, which corresponds to M/\lj$\sim$2.4--3, assuming a B-J colour for
this galaxy of 3.1 (from \ned), which is larger than our best-fitting M/\lj.
However, the blue band they adopted is more sensitive to any contamination by a young stellar population,
so that the stellar mass may not follow B-band light as faithfully as the J-band light. Furthermore,
the simplifications in their modelling and the use of a logarithmic (rather than an NFW)
dark matter halo potential (which can affect the best-fitting M/L; \S~\ref{sect_syserr_dm}) make 
the implications unclear. 
More general axisymmetric, orbit-based models that incorporate a 
dark matter halo (also using the logarithmic potential) 
have been constructed for NGC\thin 4649 by \citet{shen09a}, who obtained
M/L$_V = 8.0\pm 0.9$\msun\ (roughly $\sim$70\%\ higher than our measurement, when correcting
for the galaxy colour).  The authors were not able, however, to explore fully a range of 
inclination angles, which could affect the recovered M/L$_V$ ratio \citep[\eg][]{gavazzi05a,krajnovic05a}.
Still, we discuss the possible implications of this discrepancy in \S~\ref{sect_he}.


\subsection{Hydrostatic equilibrium} \label{sect_he}
{Our results suggest that the hot ISM of these galaxies must be close to hydrostatic.
Strong, highly localized deviations from this approximation would likely
 manifest themselves as significant
residuals in our fits to the density and temperature profiles, which are not 
generally seen (Fig~\ref{fig_profiles}). More global deviations would result in
a systematically mis-estimated pressure and consequently a bias in the measured
enclosed mass, which is not borne out by the 
good agreement between the measured stellar M/\lj\ ratios and the predictions
of SSP models for a reasonable IMF (\S~\ref{sect_mass_to_light}).
The agreement we find between our \mbh\ measurements  and those
from different techniques supports the hydrostatic approximation 
extending into the central parts of the 
galaxies (albeit less emphatically on account of the larger error-bars).

In practice
the selection of galaxies which are not significantly morphologically disturbed
in the X-ray, at least over the scales under scrutiny (excepting possible cavities, 
which we excluded from our analysis), eliminated galaxies in which hydrostatic
equilibrium is most likely perturbed.
Still, deviations from hydrostatic equilibrium could persist which do not obviously
disturb the image, due to bulk or turbulent 
gas flows, magnetic fields or cosmic rays. Each of these effects
will likely contribute nonthermal pressure support, 
causing X-ray mass techniques to {\em underestimate} the true mass
\citep[\eg][]{nagai07a,zappacosta06a,churazov08a,fang09a}. 
If such nonthermal pressure were prevalent
in these galaxies it must be compensated for by errors in the SSP models
themselves, which must underestimate the mass by almost the same amount. As 
discussed in  \S~\ref{sect_mass_to_light}, at least for two of the galaxies,
NGC\thin 4261 and NGC\thin 4649, this cannot be achieved simply by allowing  a 
complex star-formation history, and requires a modification of the IMF from the 
\citeauthor{kroupa01a} or \citeauthor{chabrier03a} form. Adiabatic contraction, or 
a similar effect, would also increase the gravitating mass in the central part of the 
galaxy and thus could  offset the effects of such putative nonthermal pressure without requiring
a modification  of the IMF. In either case, however,  the magnitude of the effect
would have to be, serendipitously, balanced with the nonthermal pressure.

Taking at face value the M/L ratios inferred from the PEGASE SSP models
that use a \citeauthor{kroupa01a} IMF (which, of all the tested models, 
agree best with the measurements), one can 
place an {\em upper limit} on the
average global fraction of the pressure which is nonthermal ($\alpha$)
by fitting a model of the form:
\begin{equation}
M/L_J (observed)  = (1-\alpha) M/L_J (SSP)
\end{equation}
To take into account the errors on both x and y axes, we used a procedure similar
to that outlined in \citet{nr}.
We obtained a marginally acceptable fit ($\chi^2$/dof=7.4/3) with
$\alpha<0.02$ (3-$\sigma$ confidence region),
indicating that no more than a few percent of the pressure support could be 
due to deviations from hydrostatic equilibrium. More conservatively, we estimated the 
effect of inherent errors in the models by cycling through the results for
the three different SSP codes.
While this does not account for the possibility of a complex star-formation
history, the failure to account for that effect would mean that $\alpha$ is 
{\em over-estimated} (at least for NGC\thin 4261 and NGC\thin 4649,
which provide crucial high-M/\lj\ leverage).
The largest allowed value of $\alpha$ was $\sim$0.18 (at 3--$\sigma$
significance) for \citeauthor{maraston05a}'s models.

Assuming the SSP models are correct, these constraints on nonthermal
pressure are particularly  important since two of the galaxies, 
NGC\thin 4261 and NGC\thin 4472, host active AGNs and associated
radio lobes that have carved out cavities in the ISM. The very small 
value of $\alpha$
dramatically illustrates that the mere existence of cavities
does not immediately imply that the ISM is out of hydrostatic equilibrium
 in regions
away from the cavities (which were excluded from our analysis). Such 
stringent limits on deviations from hydrostatic equilibrium are consistent 
with the similar $\sim$10--20\%\ limit reported by 
\citet{churazov08a} for two galaxies which are manifestly more 
disturbed than our sample. Theoretical structure formation models 
which produce morphologically relaxed-looking objects similarly suggest
no more than $\sim$25\%\ nonthermal pressure due to turbulence
\citep{tsai94a,evrard96a,nagai07a}.
 While we would not
advocate the routine use of objects as morphologically disturbed as 
M\thin 84 \citep{finoguenov01} or NGC\thin 4636 \citep{jones02a} for mass analysis 
\citep[for some of the resulting issues see][]{brighenti97a}, 
taken together with our work, these results illustrate that {\em there is little
evidence, either observational or theoretical, that
mild morphological disturbances in the X-ray image 
translate into errors on the derived mass profile that are much larger than the other
systematic errors we assessed in this paper}.
This point speaks to the recent criticisms of \citet{diehl07a}, who argued that
the lack of a correlation between the ellipticities of the X-ray and optical isophotes
for a galaxy sample indicated the ubiquity of strong deviations from hydrostatic equilibrium.
In part the lack of such a correlation was driven by the inclusion
of a significant number of objects with X-ray features carved by AGN 
ejecta. As discussed above, while these disturb the X-ray morphologies, 
they may not lead to substantial violations of hydrostatic equilibrium.
Furthermore, as pointed out by \citet{brighenti09a}, 
highly subsonic turbulence and gas rotation
(far below the level at which the enclosed mass inferred from hydrostatic techniques
is significantly biased) can substantially affect the X-ray ellipticity
at the small scales studied by \citet{diehl07a}, thus
seriously undermining the premise of their study.

Notwithstanding the small amount of nonthermal pressure implied by our results, 
larger, albeit still modest, discrepancies have been reported between the mass inferred from stellar
dynamics and hydrostatic X-ray methods at large scales in some systems 
\citep{shen09a,romanowsky08a}. It is unclear to what extent the discrepancies
arise due to deviations from hydrostatic equilibrium in the studied systems
and to what extent systematic errors in the stellar modelling play a role
(especially those due to triaxiality and the uncertain inclination of the galaxies).
\citeauthor{shen09a} argue that the stellar dynamics in one system over-estimates the 
mass inferred from X-rays, allowing room for nonthermal pressure but only if the IMF
is closer to a Salpeter than a \citeauthor{kroupa01a} form. Conversely, 
\citeauthor{romanowsky08a} argue that the mass inferred from
globular cluster dynamics in another system, NGC\thin 1407,
{\em underestimates} the X-ray measurement. This is more difficult to reconcile with
deviations from hydrostatic equilibrium, unless the gas is globally outflowing within
the central $\sim$20~kpc while maintaining a high X-ray luminosity and smooth-looking
X-ray isophotes \citepalias{humphrey06a}. Nevertheless, these studies can be used to place, entirely independently of our discussion of the SSP models,
strict upper limits on the systematic errors expected for general
use of the X-ray method. This would imply the bias on \mbh\ is no larger than
$\pm$0.3~dex but, as discussed above, we expect the actual systematic
errors due to the hydrostatic approximation to be far smaller in the present
paper. Ultimately direct observational constraints on the 
ubiquity of turbulent nonthermal pressure in the ISM of early-type galaxies and galaxy 
clusters should become possible with the advent of high-resolution non-dispersive X-ray 
spectroscopy, as enabled by \astroh\ \citep{takahashi08a} and the International
X-ray Observatory \citep[\ixo:][]{white09b}.}
\subsection{Traditional analysis} \label{sect_nonparametric}
{In this paper, we derived constraints on the mass profiles of a sample of early-type
galaxies by a ``forward fitting'' approach, wherein we parameterized the mass and
entropy profiles and used them to derive density and temperature profile models to fit to the data.
In this section, we discuss an alternative, more traditional approach that
has widely been used to model the mass profiles of galaxy clusters.
We demonstrate
that it gives consistent results with our preferred approach, albeit with larger
systematic uncertainties. This method} involves 
first fitting parameterized models to the density and temperature profiles,
and then inserting these models into the equation of hydrostatic 
equilibrium,
\begin{equation}
M(<r) = r\frac{kT}{G\mu m_H}\left( -\frac{d \ln \rho}{d \ln r} - \frac{d \ln T}{d \ln r}\right) \label{eqn_hydrostatic}
\end{equation}
to infer the mass \citep[\eg][]{lewis03a}. Here G is the universal gravitational 
constant, $\mu$ is the mean molecular weight of the gas ($\sim$0.62) 
and $m_H$ is the mass of hydrogen. 

The problem with the traditional approach is that the 
mass is primarily determined by the {\em derivatives} of arbitrary functions
fitted to the {\em binned} density and temperature data. 
In principle there exists 
an infinite family of parameterized models,
each having an arbitrary number of maxima and minima 
in any given data-bin, that, when averaged over the bin, 
equally well fit the data but which correspond to different mass profiles (many of 
which may be unphysical). The fine detail of the true density and temperature profiles can be disguised by the binning, and may not
be produced by an ad hoc, smooth parameterized model. This problem is particularly serious
when the temperature and density data-points
have large errors since the data may appear adequately fitted by a very smooth
model that does not, for any set of its parameters, resemble the true profile for the 
inferred mass distribution.  In such circumstances, the range of
mass profiles statistically allowed may not even bracket the true mass distribution.
These systematic errors \citep[\eg][]{gastaldello07a} were the prime motivation for 
the forward-fitting method we introduced in \citetalias{humphrey06a} and \citetalias{humphrey08a} 
(and used in the present work), since those methods enforce a physical mass distribution.
Furthermore, the models employed in the present paper only involve adopting an ad hoc
parameterization for the entropy profile; hydrostatic equilibrium requires 
this to rise monotonically, thus precluding the erratically rising and falling shapes 
which can
confound attempts to parameterize the binned temperature and density profiles.

Despite these concerns, the traditional method can 
be of use if the temperature and density profiles
are sufficiently close to a smooth, parameterized form to provide
a (frequentist) 
consistency check on our more formal (Bayesian) methods. 
We therefore selected appropriate, albeit 
arbitrary, smooth models to fit the density and temperature data
(described in detail in the Appendix), and hence
derived mass ``data-points'' from Eqn~\ref{eqn_hydrostatic}, 
as described in \citet{gastaldello07a}.
The choice of model was not only dictated by its ability to 
match the temperature or density data, 
but also to generate a mass
model which broadly resembles the expected form.
The best-fitting temperature and density models
are shown in Fig~\ref{fig_profiles}, and the derived mass ``data-points''
are shown in Fig~\ref{fig_mass_profiles}. 
Clearly, there
is excellent agreement between the forward-fitting and traditional 
profiles, giving us confidence that the mass models adopted in \S~\ref{sect_results} are not seriously in error.
Notable deviations between
the two methods are only seen in NGC\thin 4261, outside $\sim$15~kpc, which likely reflects the 
limitations of our ad hoc density and temperature parameterizations.

{To quantify the ability of the traditional method to place constraints on \mbh,
we fitted the set of mass ``data points'' shown in 
Fig~\ref{fig_mass_profiles} 
 with a model
comprising the dark matter halo, the stellar mass and the black hole 
(the gas mass is negligible
over the radial span of the data; Fig~\ref{fig_mass_profiles}). We obtained a good 
fit for all three galaxies ($\chi^2$/dof = 5.1/4, 8.1/7 and 13.9/8 for NGC\thin 1332, 
NGC\thin 4261 and NGC\thin 4472, respectively), and measured \mbh=$(1.00\pm0.39)\times 10^9$\msun,
$(1.2\pm0.2)\times 10^9$\msun\ and $(0.32\pm0.23)\times 10^9$\msun, respectively,
while the stellar M/\lj\ is $1.05\pm0.08$\msunlsun, $1.57\pm0.03$\msunlsun\ and $1.40\pm 0.07$\msunlsun,
respectively. Although these measurements are broadly similar to those obtained from
the forward-fitting method, it is clear that the error bars do not uniformly overlap 
between the two approaches. This is because the small error-bars obtained from the 
traditional method actually mask the much  larger systematic uncertainties discussed above.
To gain an insight into the magnitude of these effects, we cycled through 
each of the density and temperature parameterizations given in the Appendix,
and recomputed the mass profiles, provided these models were able to fit the 
data adequately well. This cycling 
had a dramatic effect on the mass profiles, with some
models even implying unphysical negative masses for some parts of the radial range.
 Fitting the mass models directly to the derived mass data points
for each parameterization, we found that 
\mbh\ varied from its ``best-fitting'' value by $10^9$\msun, 1.2$\times 10^9$\msun\ and 
$0.7\times 10^9$\msun, respectively, for NGC\thin 1332, NGC\thin 4261 and 
NGC\thin 4472. Similarly, 
M/\lj\ varied from its ``best-fitting'' value by 0.16\msunlsun, 1.6\msunlsun\ and 
$0.35$\msunlsun, respectively. These large changes indicate that systematic, rather than
statistical, errors dominate the traditional measurements for these galaxies.}
\subsection{Bondi Accretion Rates}
\begin{deluxetable}{lll}
\centering
\tablecaption{Bondi Accretion Rates\label{table_bondi_rate}}
\tablehead{\colhead{Galaxy} & \colhead{$r_{A}$}& \colhead{log$_{10}$\mbondi} \\
\colhead{} & (pc) & \colhead{([$M_\odot\ yr^{-1}$])}
}
\startdata
NGC\thin 1332 & $45^{+36}_{-25}$ & $-1.94^{+0.66}_{-0.76}$\\
NGC\thin 4261 & $24^{+15}_{-13}$ & $-2.26^{+0.56}_{-0.57}$\\
NGC\thin 4472 & $33^{+31}_{-17}$ & $-2.3^{+0.66}_{-0.65}$\\
NGC\thin 4649 & $165^{+33}_{-54}$ & $-1.41^{+0.29}_{-0.33}$\\
\enddata
\tablecomments{Accretion radii and Bondi accretion rates for each galaxy, computed from our fitted
density and temperature models (see text). Error-bars correspond to 90\%\ confidence limits.}
\end{deluxetable}
Given the reliability of our three-dimensional gas temperature and density models
(as inferred from the accuracy of our mass fits),
we were able to estimate the Bondi accretion rate for each galaxy 
\citep{dimatteo03a,pellegrini05a,allen06a} without relying on the
ad hoc extrapolations sometimes employed. 
{For adiabatic, spherically symmetric accretion from infinity onto an SMBH,
the accretion rate is given by \citep{frank92a}:
\begin{eqnarray}
\dot{M}_{bondi} & = &  \pi G^2 M_{BH}^2 \frac{\rho_g}{c_s^3} = 
\frac{\pi G^2 M_{BH}^2 \left(\frac{3}{5}\mu m_H\right)^{\frac{3}{2}}}
{\left( \rho_g^{-\frac{2}{3}} kT \right)^{\frac{3}{2}}} \nonumber \\
& = & 0.0126\left(\frac{M_{BH}}{10^9 M_\odot}\right)^2 s^{-\frac{3}{2}} M_\odot\ yr^{-1}
\label{eqn_bondi}
\end{eqnarray}
where $c_s$ is the adiabatic sound speed of the gas and s is the entropy proxy (in keV $cm^{2}$),
which is conserved by an adiabatic flow. If we 
assume that an adiabatic Bondi flow exists inwards of some ``transition
radius'', one can then compute the accretion rate from
\mbh\ and the entropy measured at any point in the flow (and without needing
to know explicitly the transition radius).  Clearly this is only a crude estimate of
the accretion rate since the calculation ignores radiative losses in the flow
and neglects properly accounting for the boundary conditions at the
transition radius. Furthermore, 
our models assume that hydrostatic equilibrium holds  at all radii
so that  if 
such a Bondi flow exists it could introduce systematic errors into our 
calculation of \mbh\ from the X-rays. Nevertheless, as discussed in \S~\ref{sect_mbh},
such errors do not appear large.  

\citet{allen06a} assumed that the 
transition to a Bondi flow occurs at the ``accretion radius'', 
$r_A$ ($=2G M_{BH}/c_s^2(\infty)$,  where
$c_s(\infty)$ is the sound speed far from the region of influence of the SMBH: \citealt{frank92a}),
that is, the point at which the gas flow becomes supersonic in a conventional Bondi flow.
For only one of our  galaxies  can we actually resolve $r_A$ (for NGC\thin 4649 it is $\sim 160$pc
$\simeq$2\arcsec) and the entropy profile is, indeed, very close to adiabatic 
at this scale \citepalias{humphrey08a}. Still, the best-fitting $s_0$ 
for the other galaxies is non-zero, suggesting  a similar flattening of the 
entropy profiles at the smallest (unresolved) scales. We therefore found it convenient 
to use the asymptotic ($r\rightarrow0$) value of $s$ to
evaluate Eqn~\ref{eqn_bondi}.  We list the most probable, marginalized values and 
90\%\ confidence regions for \mbondi\
in Table~\ref{table_bondi_rate}. \citet{allen06a} also evaluated \mbondi\
for NGC\thin 4472 and their reported value agrees, within errors, with our result.

For a sample of galaxies, \citet{allen06a} found that  \mbondi\ correlates tightly with 
an estimate of the total AGN power ejected in the jets.
This jet power was estimated under the assumption that the thermal energy of the ISM
is increased by the work done in creating cavities with a relativistic gas, so that the
mean power in the jet can be obtained from the gas pressure, the cavity size and an 
estimate of the time to create it.  \citet{mathews08a} argued, however, that 
cavities created with cosmic rays do not heat the gas in this way and should, instead drive
large-scale circulation of the gas. In this case, the jet power calculation 
is likely a very inaccurate way to estimate the energy injected into the ISM by the AGN.
Despite these concerns, the tight correlation found by \citeauthor{allen06a} is intriguing,
and it is interesting to investigate whether the galaxies in our sample are consistent
with it. 

The values of \mbondi\ in Table~\ref{table_bondi_rate} are comparable with the medium-to-high
accretion rate systems reported by \citeauthor{allen06a}, implying that the jet power should be
considerable for all of these systems if they lie on the same relation.
While there is clear evidence of AGN-blown 
cavities in NGC\thin 4261 and NGC\thin 4472, the situation is much less clear for the other two
systems, which have {\em higher} \mbondi. 
The X-ray image of NGC\thin 1332 is smooth and relaxed and exhibits little or
no evidence of X-ray cavities (Fig~\ref{fig_images}). Nonetheless, the 
modest X-ray surface brightness of this object, coupled with the lack of 
obvious radio lobes (given the weak emission in the NVSS image) make the identification
of small, low-contrast X-ray depressions challenging. More significant, however,
is the absence of cavities in a deep \chandra\ observation of the X-ray bright NGC\thin 4649
\citepalias{humphrey08a}. Based on much shallower data,
\citet{shurkin07a} claimed to find small X-ray cavities (which we did not confirm), but
even taking these at face value, 
the implied jet power estimate is only $\sim1.3\times 10^{42}$\ergps.
For this system, therefore, the jet power is, at most, comparable to the lowest values
found by \citeauthor{allen06a}, while \mbondi\ compares to their  highest Bondi rates,
in stark discrepancy with their correlation.
The existence of galaxies with high \mbondi\ but no cavities suggests that the 
\citeauthor{allen06a} correlation takes time to be established; we may be observing
NGC\thin 4649 and NGC\thin 1332 at an unusual time, while the AGN is just turning on
and the radio lobes are beginning to inflate.}
\section{Conclusions}
We have presented new hydrostatic, X-ray models for the centres of three early-type galaxies
observed with \chandra. Combined with our recent study of 
the elliptical galaxy NGC\thin 4649 \citepalias{humphrey08a},
we found, in summary:
\begin{enumerate}
\item The black hole masses measured by our method are in agreement with those
obtained by stellar or gas dynamics techniques. This provides support not only for 
our new X-ray approach, but also for the accuracy of the dynamical methods.
\item Like in  NGC\thin 4649, the black hole in NGC\thin 4472 is unambiguously required by our mass
models; even with a conservative, flat prior for \mbh, it is detected at the $\sim$2.3--$\sigma$ level.
\item Accurate stellar stellar M/\lj\ ratios require careful modelling of the stellar mass component;
with our detailed models, we obtained stellar M/\lj\ ratios in
agreement with the predictions of {single-burst 
stellar population synthesis (SSP) models computed
for a \citeauthor{kroupa01a} or \citeauthor{chabrier03a} IMF.}
This leaves little room for the steepening of the central dark matter 
density profile predicted by models of ``adiabatic contraction'', unless it exists in
a conspiracy with nonthermal pressure.
\item {Taking the SSP models at face value, this agreement suggests the gas
is very close to hydrostatic and nonthermal 
pressure can provide no more than $\sim$10--20\%\ of the total support,
unless there is a conspiracy between the shape of the IMF and nonthermal pressure.}
\item {The two galaxies with the highest Bondi accretion rates exhibit little or no
evidence of X-ray cavities, suggesting that the \citet{allen06a} correlation with
the AGN jet power takes time to be established.}
\end{enumerate}
\acknowledgements
We would like to thank Greg Martinez, Aaron Barth, H\'el\`ene Flohic and 
Sebastian Heinz for helpful discussions.
This research has made use of data obtained from the High Energy Astrophysics
Science Archive Research Center (HEASARC), provided by NASA's Goddard Space
Flight Center.
Some of the data presented in this paper were obtained from the Multimission Archive at
the Space Telescope Science Institute ({MAST}).
STScI is operated by the Association of Universities for Research in Astronomy, Inc.,
under NASA contract NAS5-26555.
This research has also made use of the
NASA/IPAC Extragalactic Database (\ned)
which is operated by the Jet Propulsion Laboratory, California Institute of
Technology, under contract with NASA, and the HyperLEDA database
(http://leda.univ-lyon1.fr).
Partial support for this work was provided by NASA under 
grant NNG04GE76G issued through the Office of Space Sciences Long-Term
Space Astrophysics Program. Partial support was also provided 
by NASA through Chandra Award Number G07-8083X issued by the Chandra X-Ray Center, which is 
operated by the Smithsonian Astrophysical Observatory for and on behalf of NASA. 

\appendix
\section{Traditional analysis models}
\begin{deluxetable*}{llrrrrrr}
\tabletypesize{\scriptsize}
\centering
\tablecaption{Traditional analysis models\label{table_traditional_models}}
\tablehead{\colhead{Galaxy} & \colhead{Model} & \colhead{$A_1$} & 
  \colhead{$A_2$} & \colhead{$R_c$} & \colhead{$R_{c2}$} & \colhead{$\alpha$} & \colhead{$\beta$}  \\
\colhead{} & \colhead{} & \colhead{($g\ cm^{-3}$; keV)} & \colhead{($g\ cm^{-3}$; keV)} & \colhead{(kpc)} & \colhead{(kpc)}
}
\startdata
\multicolumn{8}{c}{Density models} \\ \hline
NGC\thin 1332 & Eqn~\ref{eqn_cusp_beta} & $ (4.6\pm 0.7 ) \times 10^{-25} $ & \ldots & $ 0.25$  & \ldots & $ 0.98\pm 0.07 $ & $ 0.47\pm 0.008 $ \\
NGC\thin 4261 & Eqn~\ref{eqn_double_beta} & $ (1.6\pm 0.1 ) \times 10^{-24} $ & $ (2.1\pm 0.4 ) \times 10^{-27} $ & $ 0.198\pm 0.008 $ & $ 42\pm 8 $ &  \ldots  & $ 0.543\pm 0.007 $ \\
NGC\thin 4472 & Eqn~\ref{eqn_double_beta} & $ (7.9\pm 1 ) \times 10^{-25} $ & $ (6.4\pm 3.4 ) \times 10^{-27} $ & $ 0.25\pm 0.02 $ & $ 15\pm 18 $ &  \ldots  & $ 0.48\pm 0.02 $ \\
\hline
\multicolumn{8}{c}{kT models} \\ \hline
NGC\thin 1332 & Eqn~\ref{eqn_falling_temp1} & $ 0.39\pm 0.02 $ & $ 0.3\pm 0.06 $ & $ 2.1\pm 1.4 $  & \ldots & $ 5\pm 0.5 $ \\
NGC\thin 4261 & Eqn~\ref{eqn_rising_temp2} & $ 0.609\pm 0.006 $ & $ 0.75\pm 0.06 $ & $ 6.4\pm 0.6 $  & \ldots & $ 2.8\pm 0.4 $ \\
NGC\thin 4472 & Eqn~\ref{eqn_rising_temp2} & $ 0.65\pm 0.01 $ & $ 0.42\pm 0.05 $ & $ 4\pm 0.5 $  & \ldots & $ 2\pm 1 $ \\
\enddata
\tablecomments{Best-fitting results for the arbitrary, parameterized models
to the temperature and density profiles used in our ``traditional'' 
analysis (\S~\ref{sect_nonparametric}). For each galaxy, we identify
which of the functional forms given in the Appendix are used (``Model'')
and, for each model parameter, 
the best-fitting values and 1-$\sigma$ errors.}
\end{deluxetable*}
We here outline the explicit models used to parameterize the density and temperature 
profiles in our ``traditional analysis''  (\S~\ref{sect_nonparametric}). These models
are entirely {\em ad hoc} and were motivated only insofar as they reasonably capture
the global shape of the appropriate profile. In practice, the particular combination of density and 
temperature model preferred to fit the \chandra\ data of each galaxy (which is given in 
Table~\ref{table_traditional_models}) was chosen 
as that  which most accurately reproduces the mass profile derived in \S~\ref{sect_results}. 

To fit the density profile we used either a ``cusped beta model'',
\begin{equation}
\rho = A_1 \left( \frac{R}{R_c}\right)^{-\alpha} 
\left( 1 + \frac{R^2}{R_c^2} \right)^{-\frac{3}{2}\beta+\frac{1}{2}\alpha}
\label{eqn_cusp_beta}
\end{equation}
or a ``double beta model'',
\begin{equation}
\rho = \sqrt{A_1^2 \left( 1+\frac{R^2}{R_{c}^2} \right)^{-3 \beta} +
A_2^2
\left(  1+\frac{R^2}{R_{c2}^2} \right)^{-3 \beta} } \label{eqn_double_beta}
\end{equation}
and to fit the temperature data we adopted one of the following arbitrary models:
\begin{equation}
kT = A_1 +A_2 \left( 1+\frac{R}{R_c} \right)^{-\alpha} \label{eqn_falling_temp1}
\end{equation}

\begin{equation}
kT = A_1 +  A_2 \left( 1+ \left[ \frac{R}{R_c} \right]^{-\alpha} \right)^{-1}
\label{eqn_rising_temp2}
\end{equation}

\begin{equation}
kT = A_1 \left( \frac{R}{R_c} \right)^{\alpha} 
\exp \left( -\frac{R}{R_c} \right) +
A_2 \left( \frac{R}{R_c} \right)^{\beta} 
\left( 1 - \exp \left[ -\frac{R}{R_c} \right] \right)
\label{eqn_pow2expcut2}
\end{equation}
The best-fitting parameters for each galaxy are given
in Table~\ref{table_traditional_models}.

\bibliographystyle{apj_hyper}
\bibliography{paper_bibliography.bib}

\end{document}